\documentclass[aps,prd,twocolumn,notitlepage,nofootinbib,nobibnotes,superscriptaddress,amsmath,amssymb,preprintnumbers]{revtex4-1}

\usepackage{subcaption}
\usepackage{graphicx} 
\usepackage{dcolumn}
\usepackage{bm}        % for math
\usepackage{amssymb}   % for math
\usepackage{adjustbox}  
\usepackage{amsmath,array}
\usepackage{comment}
\usepackage{natbib}
\usepackage{MnSymbol}
\usepackage[hidelinks]{hyperref}
\usepackage{braket}
\usepackage     [utf8]                  {inputenc}
\usepackage     [T1]                    {fontenc}
\usepackage     [english]               {babel}
\usepackage{epigraph}
\usepackage{etoolbox}
\usepackage{ dsfont }
\usepackage{physics}
\usepackage{xcolor}
\usepackage{siunitx}
\makeatletter
\patchcmd{\epigraph}{\@epitext{#1}}{\itshape\@epitext{#1}}{}{}
\newcommand*\eqsize{%
\@setfontsize\mysize{9.0}{9.0}%
    }
    \usepackage{multirow}
\makeatother

\usepackage{xfrac}
\usepackage{amsmath}
\usepackage{esdiff} % derivatives
\usepackage{commath} % math macros
\usepackage{bbm} % blackboard style symbols
\usepackage{braket}
\usepackage{slashed}
\newcommand{\xT}{\mathbf{x}}

\newcommand{\sT}{\mathbf{s}}
\newcommand{\pT}{\mathbf{p}}
\newcommand{\qT}{\mathbf{q}}
\newcommand{\kT}{\mathbf{k}}

\newcommand{\aS}{\alpha_S}

\newcommand{\Dipper}{\textsc{McDipper}}
\newcommand{\trento}{T\raisebox{-.5ex}{R}ENTo}
\newcommand{\kompost}{K{\o}MP{\o}ST}

\definecolor{oscarC}{RGB}{22, 156, 172}

\graphicspath{{figures/}{images/}}

\begin{document}
\date{\today}

\title{Effects of sub-nucleonic fluctuations on the longitudinal structure of heavy-ion collisions}

\author{Oscar Garcia-Montero}
%\email{garcia@physik.uni-bielefeld.de}
\affiliation{Fakult\"at f\"ur Physik, Universit\"at Bielefeld, D-33615 Bielefeld, Germany}

\author{S\"oren Schlichting}
%\email{sschlichting@physik.uni-bielefeld.de}
\affiliation{Fakult\"at f\"ur Physik, Universit\"at Bielefeld, D-33615 Bielefeld, Germany}

\author{Jie Zhu}
\email{jzhu@physik.uni-bielefeld.de}
\affiliation{Institute of Particle Physics and Key Laboratory of Quark and Lepton Physics (MOE), Central China Normal University, Wuhan, 430079, China}
\affiliation{Fakult\"at f\"ur Physik, Universit\"at Bielefeld, D-33615 Bielefeld, Germany}

\begin{abstract} 
Sub-nuclear fluctuations in the initial state of heavy-ion collisions impact not only  transverse long-range correlations of small systems, but also the creation of longitudinal structures, seen in particle detectors as longitudinal decorrelation observables. In this work, we study the emergence of long-range rapidity correlations in nuclear collisions based on the 3D resolved {\Dipper}  initial state model, and for the first time, connect it to experimental observables using the 3+1D viscous hydrodynamics framework CLVisc. We include different sources of fluctuations at the nucleon and subnucleon level and study the effects of these additional fluctuation sources on the longitudinal structure of relevant observables, such as the flow decorrelations and directed flow.

%, dominating the energy deposition at mid-rapidity. Furthermore, we add momentum and spatial fluctuations on the initial energy and  the incoming high-$x$ partons by sampling quarks -and anti quarks- from Parton distribution functions (PDFs). The energy and charge fluctuations will dominate the physics at higher rapidities, in the so-called fragmentation regions. Specially important from the strange sector, the deposition of (anti-)quark  pairs allows us to create profiles with local spatial fluctuations of baryon, charge and strangeness profiles. an input needed in the study of 3D charge diffusion in the hydrodynamical stage. 
\end{abstract}

\maketitle

\section{Introduction}
\label{sec:intro}

%An excellent facility to help us understand the interactions between elementary particles is heavy-ion collisions(HIC) which smash the nuclei into quarks and gluons However, 
Describing the complex space-time evolution of heavy-ion collisions (HICs), is a challenging task in theoretical nuclear physics~\cite{Jacak:2012dx, PHENIX:2006dpn}. Nowadays, the
standard procedure adapted for collisions at high center-of-mass (CoM) energies, the so-called Standard Model of HICs~\cite{Muller:2012zq,Muller:2013dea}, describes the dynamics in terms of multi-stage evolution models. During the first stage, the initial energy and charge deposition in the collision is followed by a short epoch of far-from-equilibrium dynamics, before the Quark-Gluon Plasma (QGP) relaxes towards equilibrium and expands hydrodynamically. Eventually, as the energy density in the fluid crosses the critical energy density of the QCD crossover, the Quark-Gluon Plasma (QGP) fluid is converted into hadronic degrees of freedom, which undergo hadronic re-scatterings and resonance decays, before falling again out of chemical and kinetic equilibrium, and flying to the detectors~\cite{Muller:2013dea, Heinz:2013wva}.  By now, the last two processes can be well described by viscous hydrodynamics~\cite{Schenke:2010nt,Wu:2021fjf, Pang:2018zzo} and hadronic transport models, the so-called hadronic afterburners~\cite{SMASH:2016zqf,Bleicher:1999xi}  respectively, while a robust theoretical description of the  initial state and pre-equilibrium dynamics still requires further efforts. 

While great progress has been achieved in the last two decades, most models for initial condition and pre-equilibrium stage are intended to describe the evolution around mid-rapidity. For example, the IP-Glasma~\cite{Schenke:2012wb, Schenke:2012hg, Mantysaari:2017cni} describes the energy deposition at mid-rapidity and {\kompost}~\cite{Kurkela:2018wud, Kurkela:2018vqr} provides the (boost-invariant) pre-equilibrium evolution of conserved charges in the central rapidity region.
Even if the boost-invariant condition in longitudinal direction works well at mid-rapidity in high energy collisions, a realistic longitudinal dynamics is urgently needed to describe the baryon-stopping~\cite{Videbaek:2009zy}, rapidity dependency of anisotropic~\cite{ALICE:2016tlx} and longitudinal decorrelation~\cite{CMS:2015xmx} observed in experiments. 
%However, a complete description including longitudinal rapidity distribution remain unresolved, whether initial condition or pre-equilibrium dynamics. 

So far, common recipes for generating longitudinal profiles in the initial instant of the collision include purely parametric models such as \trento-3D~\cite{Ke:2016jrd, Soeder:2023vdn} and transport models such as AMPT~\cite{Lin:2004en}. On the side of saturation physics, it is still a matter of current research how to extend the successful picture of the IP-Glasma towards the longitudinal direction~\cite{Matsuda:2024moa,Matsuda:2024mmr,Ipp:2017lho,Gelfand:2016yho,Schlichting:2020wrv,Ipp:2024ykh,Ipp:2021lwz}. However, multiple advances have been made in the last years including different contributions to the (3+1)D structure of the model~\cite{McDonald:2020oyf,Ipp:2024ykh,Ipp:2021lwz,Schlichting:2020wrv,Schenke:2022zkw}.
On the other hand, pre-equilibrium evolution is an important source of longitudinal structure of HIC, current studies either avoid this stage by connecting the hydrodynamic expansion to the initial conditions directly~\cite{McDonald:2020oyf} or using a free-steaming evolution~\cite{Bernhard:2019bmu} to simplify the dynamics in this stage. While an effective macroscopic description of the pre-equilibrium dynamics based on non-equilibrium linear responses theory~\cite{Kurkela:2018wud, Kurkela:2018vqr} (\kompost) has been proposed to describe the transverse dynamics at mid-rapidity in this stage, the extension of \kompost~to 3+1D dimensions is still in progress.

Interestingly, some studies have pointed out that the longitudinal behavior of final observables, such as directed flow, may in fact be very sensitive to the non-equilibrium dynamics of the early stage~\cite{Bozek:2022cjj, Bozek:2010aj}, and it should be expected that the rich interplay of transverse and longitudinal expansion can play a significant role in explaining the longitudinal structures seen in experiments.

% In fact, it has been suggested that a realistic 3+1D pre-equilibrium stage could also cause a non-boost invariant longitudinal structure to generate decorrelation, as a response to the transverse fluctuations~\cite{stephan:thesis}.\oscar{Check Ref}
% However the creation of a 3D pre-equilibrium stage which takes on accounts the QCD dynamics of the system is still work in progress. 

Before such a study of the combined initial+pre-equilibrium effects can be performed, a thorough exploration of the effects of transverse fluctuations on the longitudinal structure of heavy-ion collisions is imperative. In this work we explore the effects of initial nucleon and sub-nucleon fluctuations on the longitudinal dynamics of heavy-ion collisions. We do this by extending the {\Dipper} model\cite{Garcia-Montero:2023gex}, a state-of-the-art framework for the initial state energy and charge deposition based on the Color Glass Condensate (CGC) to incorporate event-by-event fluctuations at the sub-nucleonic level, which has been done by implementing the so-called hotspot model~\cite{Mantysaari:2016jaz, Mantysaari:2016ykx}. 

While it is well understood that fluctuations at the nucleonic scale can describe most of experimental data in heavy-ion collisions~\cite{Aguiar:2001ac, Schenke:2010rr, Schenke:2011bn, Zhao:2017yhj}, fluctuations at the sub-nucleonic level %which were proposed right after the concept of wounded nucleons to explain the particle production mechanisms
are necessary when looking at fluctuation-driven observables or when the collision system is small ~\cite{Bialas:1977en, Bialas:1978ze, Zheng:2016nxx}. Sub-nucleonic fluctuations are also important to be able to describe the anisotropy observed in p+Au, d+Au, and He3+Au collisions~\cite{PHENIX:2016cfs, Schenke:2014zha, Mantysaari:2017cni} and the production of vector mesons(e.g., $J/\Psi$)~\cite{Mantysaari:2016ykx, Mantysaari:2016jaz} in ultra-peripheral collisions (UPC). However, so far most studies have focused on the effects of sub-nucleonic fluctuations in the transverse plane and not explored the effect on longitudinal decorrelation observables.

This paper is organized as follows: 
in Sec.~\ref{Model} we present a short introduction of the framework used to simulate heavy-ion collisions,
where in Sec.~\ref{mcdipper} we first review the energy and charge deposition in {\Dipper} framework, then detail how  sub-nucleonic fluctuations are introduced. We illustrate how these new additions affect the initial energy and charge deposition as well as the multiplicity distributions. Thereafter, the interface connecting initial condition and the following hydrodynamic evolution are described in detail.
In Sec.~\ref{hydro}, we provide a brief overview of the (3+1)D viscous hydrodynamics model, CLVisc, which describes the full evolution of QGP medium including the evolution of the baryon current. The particlization of the fireball is explained in Sec.~\ref{freezeout}. 
The numerical results from our \Dipper+CLVisc integrated framework are demonstrated in Sec.~\ref{results}, including the (pseudo-)rapidity ($\eta$) dependence of charged hadron and net-proton yields, transverse momentum ($p_T$) spectra of identified particles, and the first three orders of anisotropic flow coefficients ($v_n$) as well as the longitudinal decorrelation from central collisions to peripheral collisions. The effects of sub-nucleonic fluctuations are explored in comparison to initial conditions for smooth nucleons.
In Sec.~\ref{summary}, we summarize the findings and point out possible directions for future work.

\section{\Dipper+CLVisc}
\label{Model}
\subsection{3D resolved {\Dipper}  initial state model}
\label{mcdipper}
The {\Dipper}~\cite{Garcia-Montero:2023gex} is an initial state energy and charge deposition model based on the $k_T$-factorization limit of the CGC effective field theory. In this formalism,
the energy and conserved charges can be found as moments of the single inclusive gluon and quark distributions. The results presented in this work correspond to single parton production formulas computed at leading order (LO). 
Namely, the single inclusive production of gluons arises from  radiative processes, while in the high energy limit, the single inclusive quark production contributes at the LO level via the stopping of collinear quarks inside the projectile nucleus. These quarks are deflected from their light-cone trajectories due to multiple-scattering off the constituents of the target nucleus. Generally, the energy density $(e\tau)_0$ can be computed by getting the first moment of the gluon and (sea)quark distributions,
\begin{equation}
(e\tau)_0 =\int d^2\pT~|\pT|~\left[K_g \frac{dN_{g}}{d^2\xT d^2\pT dy} + \sum_{f,\bar{f}} \frac{dN_{q_f}}{d^2\xT d^2\pT dy}\right]_{y=\eta_s}\;,
\label{eq:EnergyWithKFactor}
\end{equation}
where in accordance with \cite{Garcia-Montero:2023gex} we introduce the phenomenological normalization factor $K_g$ to account for normalization uncertainties and higher order corrections to  gluon production. %Given that the parameters needed to describe the gluon distributions of the targets are fixed to Deeply Inelastic Scattering experiments at HERA [XXX], $K_g$ stands as the only single free parameter in the saturation prescription.
Similarly, the net-charge densities $(n_f\tau)_0$ of the light flavors $f=[u,d,s]$, are the zeroth order moments of the net-quark  distribution,
\begin{equation}
(n_f\tau)_0 =\int d^2\pT~\left[\frac{dN_{q_f}}{d^2\xT d^2\pT dy} - \frac{dN_{\bar{q}_f}}{d^2\xT d^2\pT dy}\right]_{y=\eta_s} \,.
\label{eq:QuarkCharges}
\end{equation}

We will see below that at LO, without further assumptions the average strange distribution vanishes identically, i.e. $(n_s\tau)_0=0$. 

The spectrum of gluons, $\frac{dN_{g}}{d^2\xT d^2\pT dy}$, can be written as a function of transverse momentum $\pT$, momentum rapidity $y$ and transverse position $\xT$. At LO, it can be intuitively expressed as the integration  of the two momentum space dipole functions~\cite{Blaizot:2004wv,Gelis:2001da,Gelis:2001dh} of the projectile and target, in the adjoint representation, with a hard factor (see Refs.~\cite{Dumitru:2001ux,Lappi:2017skr})
% \begin{equation}
% \begin{split}
% \frac{dN_{g}}{d^2\xT d^2\pT dy}=  &\frac{g^2\,N_c}{4\pi^5 (N_c^2-1)\,\pT^2}  \int \frac{d^2\qT}{(2\pi)^2}   \frac{d^2\kT}{(2\pi)^2}~\\
% &\times~\Phi_1(x_1,\xT,\qT)~\Phi_2(x_2,\xT,\kT)\\
% &\times~(2\pi)^2\delta^{(2)}(\qT+\kT-\pT)\,\\
% \end{split}
% \label{eq:gluon_prod_w_uGDFs}
% \end{equation}
\begin{equation}
\begin{split}
\frac{dN_{g}}{d^2\xT d^2\pT dy}=  &\frac{ (N_c^2-1)}{4\pi^3 g^2 N_c}  \int \frac{d^2\qT}{(2\pi)^2}   \frac{d^2\kT}{(2\pi)^2}~\frac{\qT^2\kT^2}{\pT^2}\\
&\times~D_{1,\rm adj}(x_1,\xT,\qT)~D_{2,\rm adj}(x_2,\xT,\kT)\\
&\times~(2\pi)^2\delta^{(2)}(\qT+\kT-\pT)\,\\
\end{split}
\label{eq:gluon_prod_w_uGDFs}
\end{equation}
where  $N_c=3$ denotes the number of colors and $g$ is the strong coupling constant. 
In this limit of the CGC formalism kinematics relate $x$ and rapidity in a straightforward manner, where one can identify the light-cone momentum fraction in the projectile and target, namely $x_{1/2}=\frac{\pT}{\sqrt{s_{NN}}} e^{\pm y}$.

Similarly, the single inclusive quark spectra $\frac{dN_{q_{f}}}{d^2\xT d^2\pT dy}$, characterising the stopping of quarks and anti-quarks from both nuclei due to the interaction with the target nuclei,  can be expressed at LO as~\cite{Dumitru:2002qt,Dumitru:2005gt}
\begin{equation}
\begin{split}
\frac{dN_{q_{f}}}{d^2\xT d^2\pT dy} &= \frac{x_{1}q^{A}_{f}(x_{1},\pT^2,\xT)~D_{\rm fun}(x_2,\xT,\pT)}{(2\pi)^2} \\
&+ \frac{x_{2}q^{A}_{f}(x_{2},\pT^2,\xT)~D_{\rm fun}(x_1,\xT,\pT)}{(2\pi)^2}\,.
\end{split}
\label{eq:single_quark_density}
\end{equation}
for an (anti-)quark of flavor $f(\bar{f})$, where the kinematics of $x_{1,2}$ is identical to the single inclusive gluon production formula. Here, 
$q^{A}_{f}(x_{1/2},Q^2,\xT)$ are the collinear quark distributions for the two nuclei (see below). In this case, as quarks go through the initial gluon wave function, they pick up transverse momentum, and a color rotation phase. 
% The total effect can be now described by the fundamental dipole distribution
% \begin{equation}
%     D_{\rm fun}(x,\xT,\qT)=\frac{1}{N_c} \int_{\sT} \tr_{\rm fun}[U_{\xT+\sT/2} U^{\dagger}_{\xT-\sT/2}]~e^{\rm i \qT\cdot\sT} \\ 
% \end{equation}
In this work, we will model the color charge-nucleus interactions using the IP-Sat model, \cite{Kowalski:2006hc,Kowalski:2003hm} where the dipole is given by
\begin{equation}
  D_{\rm fun}(x,\xT,\sT)=\exp\left[-\frac{\pi^2\sT^2}{2\,Nc}\aS(\mu^2)\, xg(x,\mu^2)\, T(\xT)\right]\,.
  \label{eq:IPSatDip}
\end{equation}
The function $g(x,\mu^2)$ is the gluon distribution function (PDF) initialized at a scale $\mu_0$ via the parametrization 
\begin{equation}
xg(x,\mu^2_0)= a_g\,x^{-\lambda_g}\, (1-x)^{5.6}\,.
\end{equation}
and evolved to further scales, $\mu$, via the DGLAP equation. Additionally, in the IP-Sat model the scale at which the coupling and the gluon PDF is evaluated using $\mu^2 = \mu^2_0 + C/\sT^2$.
If one assumes that local color correlations of charges inside the nucleus are Gaussian, one obtains the relation $
D_{\rm adj}(x,\xT,\sT)=[D_{\rm fun}(x,\xT,\sT)]^{C_A/C_F}\;,
$
which we use throughout this work to obtain the adjoint dipole from the fundamental dipole given by \eqref{eq:IPSatDip}. 

We model the collinear quark/anti-quark distributions as an ensemble of uncorrelated partons, distributed in transverse space according to the local density of protons ($p$) and neutrons ($n$), as in 
\begin{equation}
    \begin{split}
    u^A(x,Q^2,\xT)&=u_{p}(x,Q^2)\; T_{p}(\xT) + u_{n}(x,Q^2)\; T_{n}(\xT)\,, \\
     d^A(x,Q^2,\xT)&= d_{p}(x,Q^2)\; T_{p}(\xT) + d_{n}(x,Q^2)\; T_{n}(\xT)\,, \\
      s^A(x,Q^2,\xT)&= s_{p}(x,Q^2)\;T_{p}(\xT) + s_{n}(x,Q^2)\;T_{n}(\xT)\,.
    \end{split}
     \label{eq:PDFS-pn}
\end{equation}
for $u,d,s$ flavor (anti-)quarks, where by $T_{p/n}(x)=\sum_{i=1}^{A}~t_{p/n}(\xT-\xT_i)$ we denote the nuclear thickness of protons and neutrons in the nucleus with atomic number $A$. In our model, we use isospin symmetry as the parton distributions of the neutron are not well constrained.  The neutron PDFs are then set to the proton PDFs after the transformation $u\leftrightarrow d$.

\subsection{Sub-nucleonic fluctuations}
We introduce sub-nucleonic fluctuations in our newest version, {\Dipper} 1.2~\cite{mcdipper2024}, where the transverse density profile of each nucleon is assumed to be the weighted superposition of $N_q$ hotspots~\cite{Mantysaari:2016jaz, Mantysaari:2016ykx},
\begin{equation}\label{eq:hotspots}
t_{p/n}\left(\mathbf{b}_{\perp}\right)=\frac{1}{N_q} \sum_{i=1}^{N_q} p_i\,t_q\left(\mathbf{b}_{\perp}-\mathbf{b}_{\perp, i}\right),
\end{equation}
and assume each hotspot presents a gaussian profile,
\begin{equation}
t_q\left(\mathbf{b}_{\perp}\right)=\frac{1}{2 \pi B_q} e^{-\mathbf{b}_{\perp}^2 /\left(2 B_q\right)}.
\end{equation}
The hotspot positions $b_{\perp,i}$ are sampled from a two-dimensional
Gaussian distribution whose width is denoted by $B_{qc}$. We set $B_{qc}=\frac{B_G-B_q}{1-1 / N_q}$ to reproduce the nucleon width $B_G$ as in \trento~\cite{Moreland:2018gsh}, where the factor in the denominator accounts for the fact that the resulting center-of-mass is shifted to the expected center of the nucleon. 

\begin{figure}
    \centering
    \includegraphics[width=\linewidth]{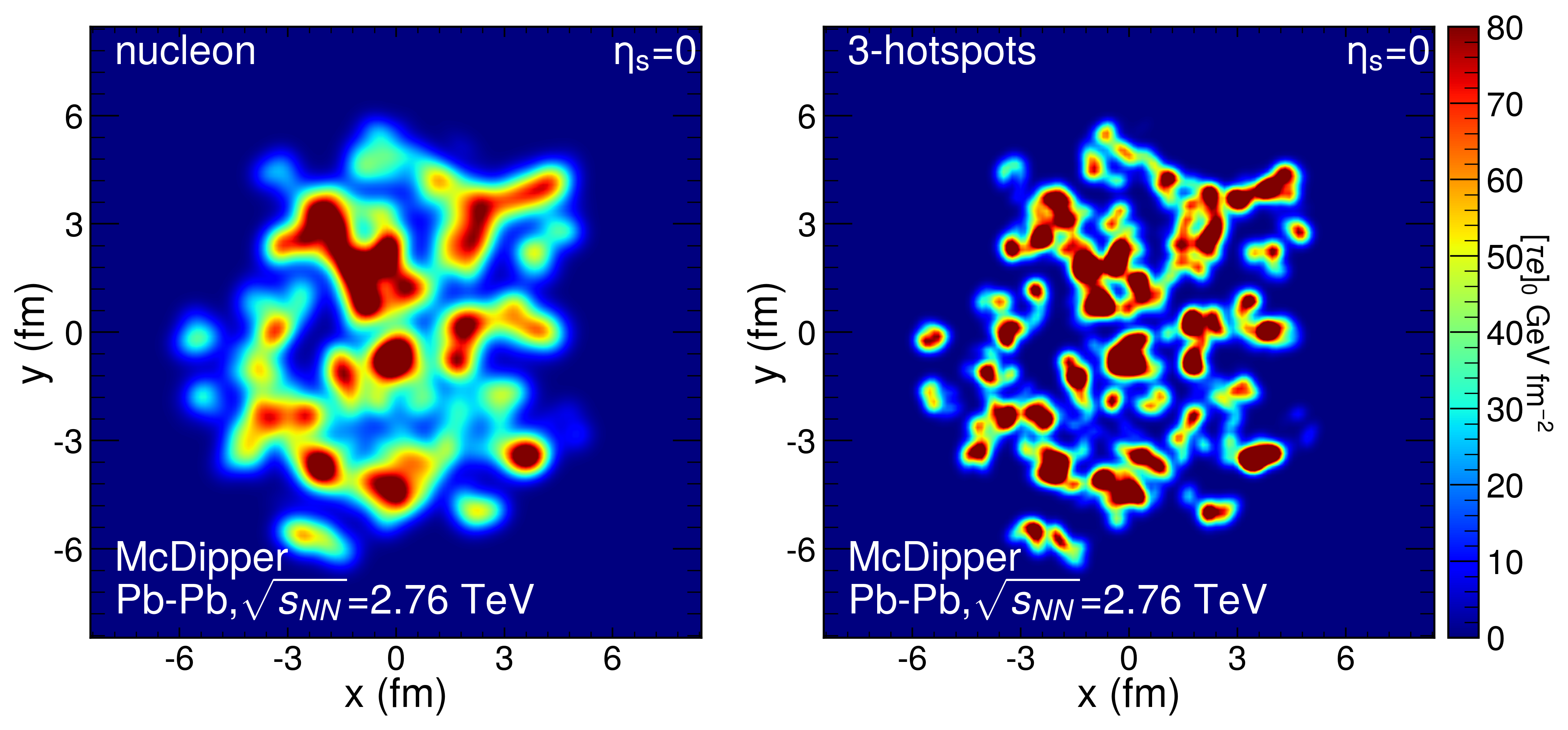}
    \includegraphics[width=\linewidth]{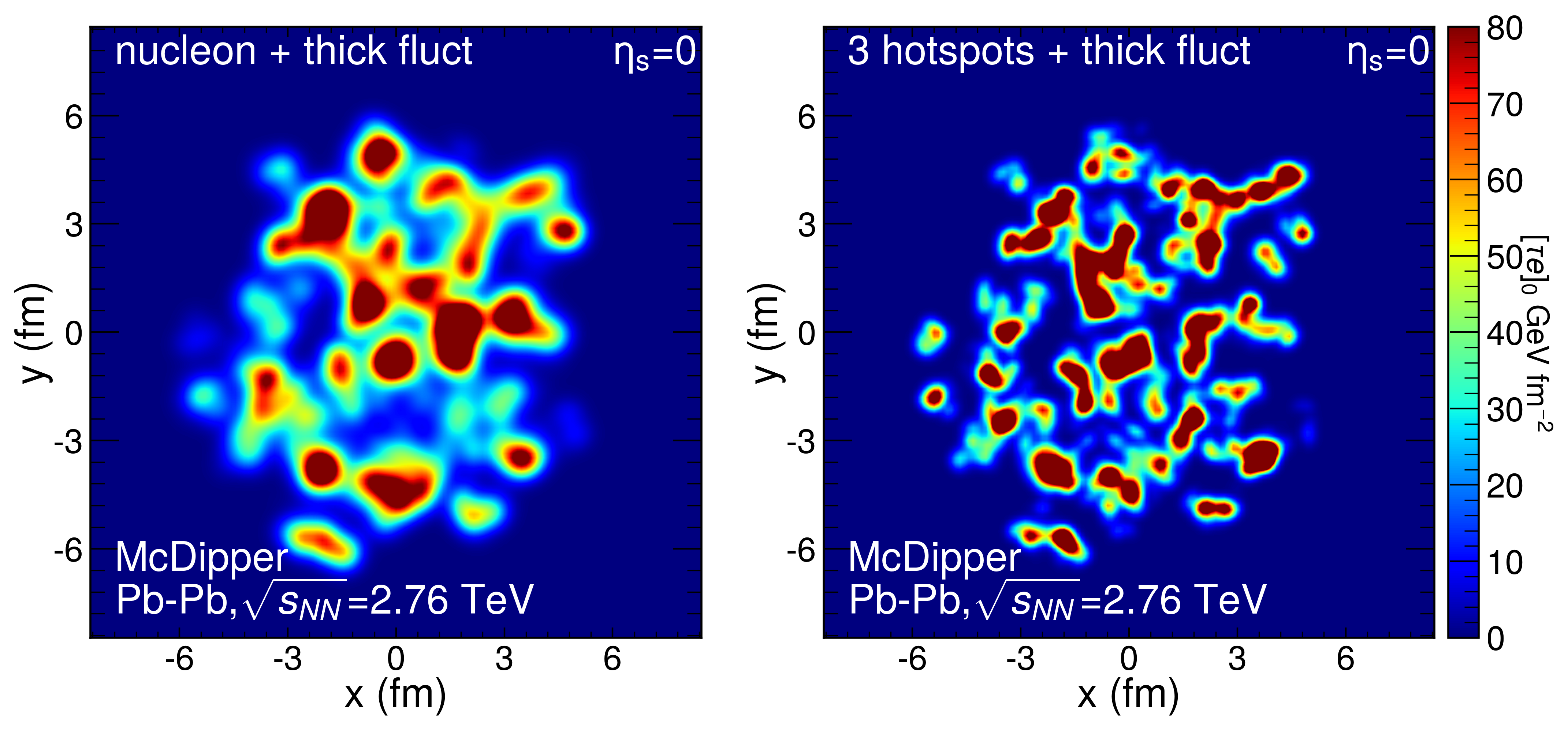}
    \caption{The initial energy density distributions of different classes at $\rm \eta_s$=0 for a sampled Pb+Pb collision at $\sqrt{s_{NN}}$=2.76 TeV and impact parameter b=3.42 fm.}
    \label{fig:ini-ed-trans}
\end{figure}

We also introduced the coefficients $p_i$ in Eq.(\ref{eq:hotspots}), which are sampled from the log-normal distribution~\cite{Mantysaari:2016jaz, Mantysaari:2016ykx} to account for the participants' thickness fluctuations
\begin{equation}
P\left(\ln p_i\right)=\frac{1}{\sqrt{2 \pi} \sigma} \exp \left[-\frac{\ln ^2 p_i}{2 \sigma^2}\right].
\label{eq:thick_fluct}
\end{equation}
Subsequently, the sampled $p_i$ are scaled by the expectation value of the distribution $E[p_i]=e^{\sigma^2/2}$ to keep the average density unchanged. The parameter $\sigma$ controls the magnitude of thickness fluctuations.
If not stated otherwise, in this work, we choose $N_q$=3, $B_q=0.04\ \rm fm^2$, $B_G=0.156\ \rm fm^2$ and we will explore two different values of $\sigma$=0.637 or 1.2.
Since we invoke sub-nucleonic degrees of freedom, we also take them into account to determine the participant status of each colliding nucleon. More precisely, the collision of two incoming nucleons is assumed to occur based on the following probability distribution
\begin{equation}
P(\vec{b})=1-\exp[-\frac{\sigma_{gg}}{N_q^2} \sum_{i,j} \frac{1}{4\pi B_q} \exp(-\frac{({\bold{b}_{\perp, i}}-{\bold{b}^\prime_{\perp, j}})^2}{4B_q})]
\end{equation}
where $\sigma_{gg}$ are tuned to reproduce the inelastic cross section in pp collisions, i.e. $\sigma_{pp}^{\rm inel}=\int d^2 \vec{b} P(\vec{b})$.

\begin{figure}
    \centering
    \includegraphics[width=\linewidth]{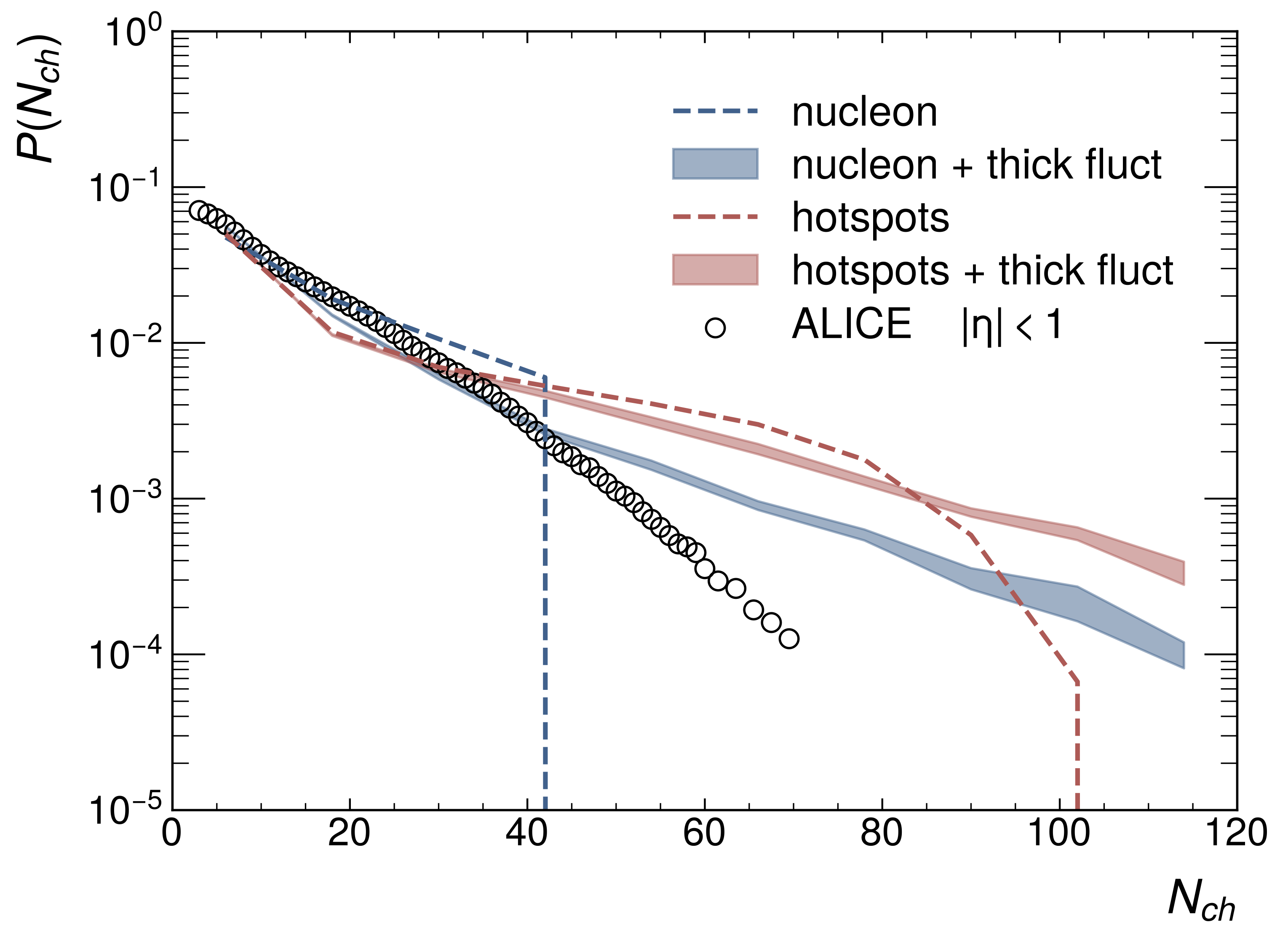}
    \caption{The effects of sub-nucleonic fluctuations on the mid-rapidity multiplicity distributions in $\mathrm{p+p}$ collisions at $\sqrt{s_{NN}}$=7 TeV.}
    \label{fig:pp-dist}
\end{figure}

To illustrate the effects of sub-nucleonic fluctuations, we present the initial energy density distributions at $\rm \eta_s$=0 for Pb-Pb collisions at $\sqrt{s_{NN}} = 2.76$ TeV in Fig.~\ref{fig:ini-ed-trans}. Four different classes of initial conditions are considered here, which are the original \Dipper\ at the nucleon level (upper left), \Dipper\ with 3 hotspots (upper right), \Dipper\ at the nucleon level with thickness fluctuations (lower left) and \Dipper\ with 3 hotspots and thickness fluctuations (lower right).
In the four classes, we run \Dipper\ for the same  configuration of the nucleons and employ the same $K_g$=2.25 for the IP-Sat saturation model.
One can see that the inclusion of 3 hotspots in each nucleon breaks the initial fireball down into smaller parts, as a result, the gradient of energy density distribution is steeper which may cause more drastic expansion and furthermore influence the total geometry. Conversely, the thickness fluctuations do not change the granularity, but  only cause a redistribution of energy density. The longitudinal energy and net-baryon density distributions at $x=0$ fm are also shown in the Fig. \ref{fig:ini-ed-long} and \ref{fig:ini-nb-long} of Appendix \ref{app-ini}.

In Fig.~\ref{fig:pp-dist}, we estimate the distribution of multiplicity in the pseudo-rapidity region [-1.0,1.0] in proton-proton collisions at $\sqrt{s_{NN}}=7$ TeV with the formula\cite{ALICE:2022imr}

\begin{equation}\label{Nch_esti}
\frac{\mathrm{d} N_{\mathrm{ch}}}{\mathrm{d} \eta} \approx
\left\langle\frac{\mathrm{d} E_{\perp}}{\mathrm{d} y}\right\rangle
\left/ \left[ \frac{\langle m\rangle}{f_{\mathrm{tot}}} \sqrt{1+a^2} J^{-1}(a, \eta)\right]\right. ,
\end{equation}
where $\langle m\rangle=(0.215 \pm 0.001) \text{GeV},\ f_{\text {tot }}=0.55 \pm 0.01,\ a=\left\langle p_{\perp}\right\rangle /\langle m\rangle\approx 1.9$ are taken from ALICE\cite{ALICE:2022imr}.
To maintain consistency, in different scenarios $K_g$ is tuned by final multiplicity in Pb-Pb collisions which will be explained in detail later. The estimated results are compared to ALICE data~\cite{ALICE:2010mty} in Fig.~\ref{fig:pp-dist} which shows that only when we include the thickness fluctuations in \Dipper\ can we generate the high multiplicity tail that appears in experimental data~\cite{ALICE:2010mty}. Discrepancies between simulation and experimental data should be attributed to the facts that $K_g$'s of different scenarios are not tuned specifically for proton-proton collisions to fit the mean value of $N_{ch}$ and that the parameters in formula (\ref{Nch_esti}) are subtracted from pp collisions at $\sqrt{s_{NN}}$=5.02 TeV. Evidently, it would also be desirable to obtain a better estimate than (\ref{Nch_esti}) of the particle production in p+p collisions, where in the absence of final state effects one could invoke a microscopic description of hadronization as e.g. in~\cite{Greif:2020rhi}, however this is beyond the scope of the present work. 

\begin{figure}[t]
    \centering
    \includegraphics[width=\linewidth]{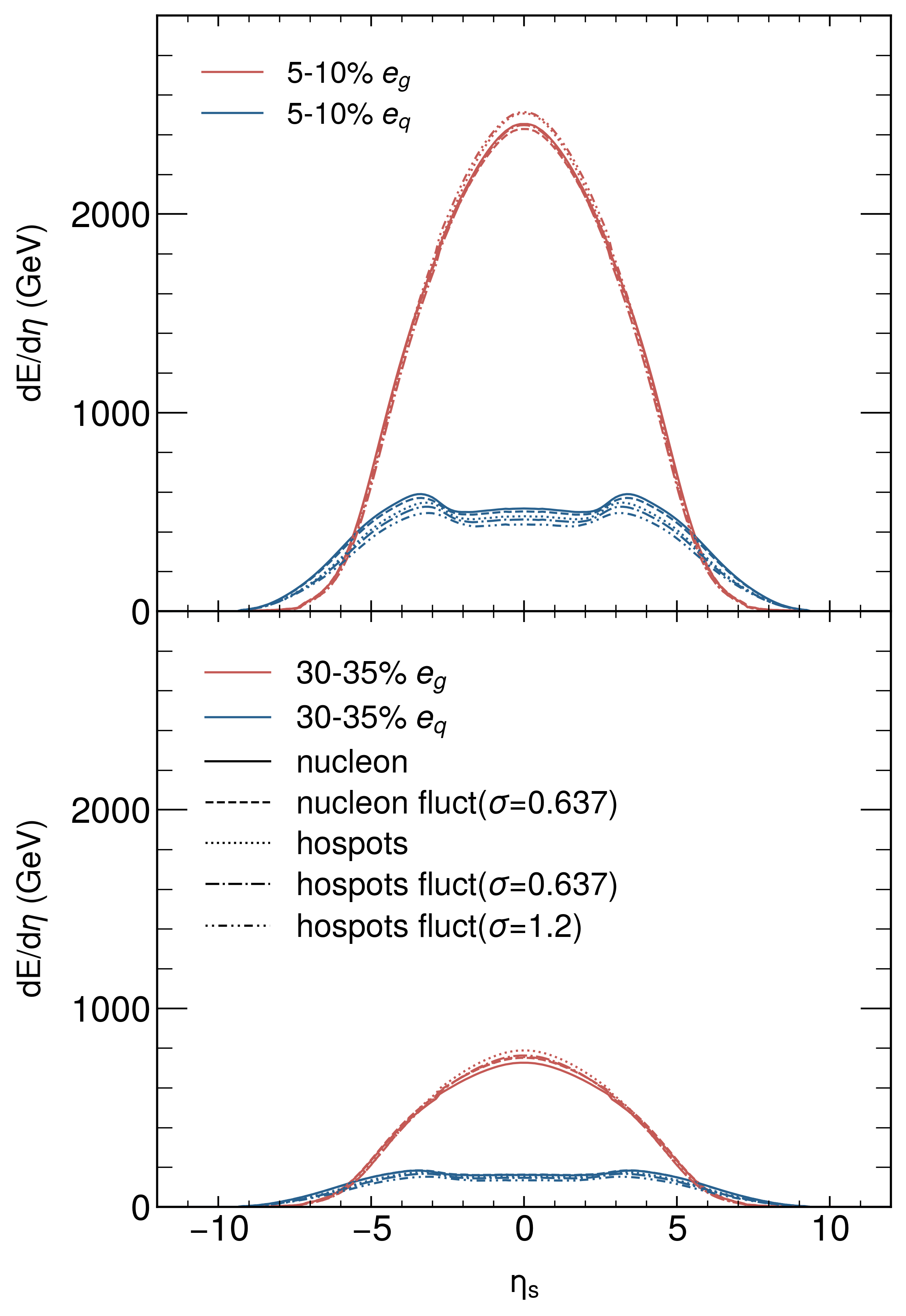}
    \caption{ The energy density along the pseudo-rapidity direction for Pb+Pb $\sqrt{s_{NN}}$ = 2.76 TeV collisions with centrality range 5-10\%, 30-35\%.}
    \label{fig:ed-vs-eta}
\end{figure}

\subsection{Initial conditions for hydrodynamic evolution}
By obtaining the energy and charge density from \Dipper, we can construct the initial conditions for the subsequent hydrodynamic evolution as follows,
\begin{equation}
\begin{aligned}
T^{\mu\nu}&=\int d^2 p_T \ p^\mu p^\nu \cdot \frac{1}{\tau p_T} \cdot \frac{dN}{d^2p_T d^2 x_T dy}\\
&=\mathrm{diag}(e,e/2,e/2,0),\\
J^\mu_B&=\int d^2 p_T\ p^\mu \cdot \frac{1}{\tau p_T} \cdot \frac{dN_B}{d^2p_T d^2 x_T dy}\\
&=(n_B,0,0,0),
\end{aligned}
\end{equation}
where $n_B=(n_u+n_d+n_s)/3$. To get the shear tensor($\pi^{\mu\nu}$) and bulk pressure($\Pi$), we decompose $T^{\mu\nu}$ and $J^\mu_B$ in Landau Frame,
\begin{equation}
\begin{gathered}
T^{\mu\nu}=\left(e+(P+\Pi)\right) u^\mu u^\nu - (P+\Pi)g^{\mu\nu}+\pi^{\mu\nu},\\
J^\mu=n u^\mu + V^\mu,
\end{gathered}
\end{equation}
such that, upon determining $u^\mu=(1,0,0,0)$ and considering the traceless property of $\pi^{\mu\nu}$, we get 
\begin{equation}
\begin{gathered}
\pi^{\mu\nu}=\mathrm{diag}(0,\frac{e}{6}, \frac{e}{6}, -\frac{e}{3\tau^2}),\\
\Pi=e/3-P, \qquad V^\mu=0\;.
\end{gathered}
\end{equation}

\subsection{Event selection and tuning}
We generate $10^4$ minimum bias initial conditions of Pb-Pb collisions at $\sqrt{s_{NN}}$=2.76 TeV  with \Dipper, and perform the centrality selection following~\cite{Giacalone:2019ldn} based on the estimator $\int \mathrm{d}^2 \mathbf{x} [\tau e(y,\mathbf{x})]^{2/3}_0$, which can be related to entropy, and therefore to  final multiplicity at mid-rapidity. Each initial state is then input into hydrodynamic evolution at an initialization time of $\tau_0$=0.6 fm.  In order to assess the effect of different kinds of fluctuations,
five sets of simulations are presented in this paper, which can be summarized as 
\begin{itemize}
\item \Dipper\ with smooth nucleons (no subnuclear fluctuations, denoted as \textit{nucleon}) 
\item \Dipper\ with 3 hotspots (denoted as \textit{hotspots})
\item \Dipper\ at the nucleon level with thickness fluctuations(denoted as \textit{nucleon fluctuations} $\sigma=0.637$)
\item \Dipper\ with 3 hotspots and thickness fluctuations of different magnitudes(denoted as \textit{hotspots fluctuations} )
\end{itemize}
\begin{figure}[t]
    \centering
    \includegraphics[width=\linewidth]{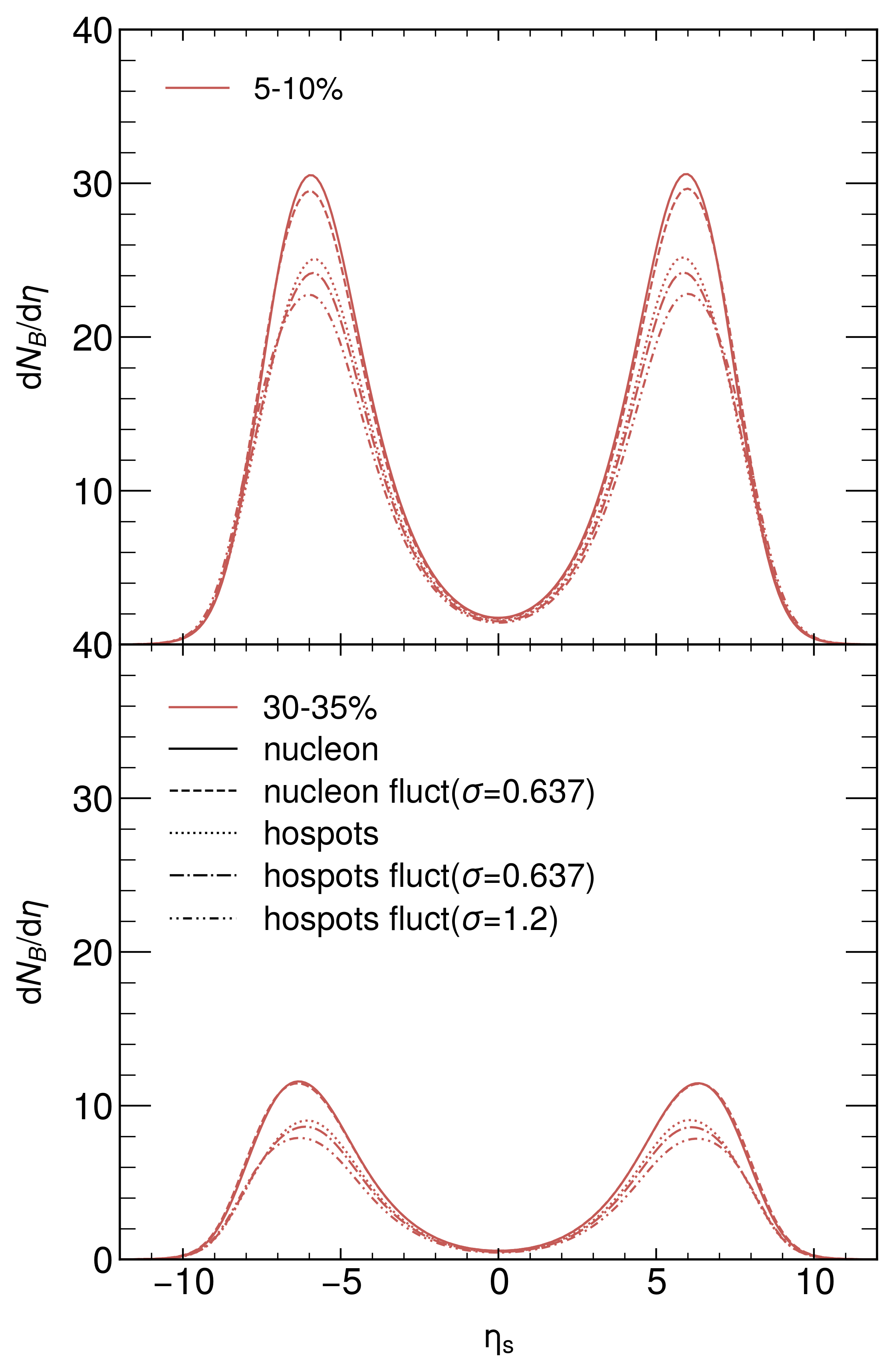}
    \caption{ The net baryon density along the pseudo-rapidity direction for Pb+Pb $\sqrt{s_{NN}}$ = 2.76 TeV collisions with centrality range 5-10\%, 30-35\%.}
    \label{fig:nb-vs-eta}
\end{figure}

where  the {\Dipper} v1.2 was used to produce all the initial conditions above~\cite{mcdipper2024}. For the fluctuations of the thickness functions, we have chosen a spread of the thickness distribution function, see Eq.~\ref{eq:thick_fluct}, of width $\sigma$=0.637 for the smooth nucleon case, and $\sigma$=0.637 or 1.2 for the hotspot case. To compare with experimental data, we tune $K_g$ for the presented scenarios
(
$K_g$=2.23 for \textit{nucleon} case,
$K_g$=2.34 for \textit{nucleon fluctuations} case with $\sigma$=0.637,
$K_g$=2.61 for \textit{hotspots} case,
$K_g$=2.71 for \textit{hotspots fluctuations} case with $\sigma$=0.637,
$K_g$=3.20 for \textit{hotspots fluctuations} case with $\sigma$=1.2
) to ensure that the charged particle multiplicity in Pb+Pb collisions matches the experimental results at mid-rapidity in 0-5\% centrality.

The initial longitudinal energy density distribution (dE/d$\rm\eta$) and net baryon number density distribution (d$N_B$/d$\rm\eta$) in centrality classes 5-10\% and 30-35\% are shown in Fig. \ref{fig:ed-vs-eta} and \ref{fig:nb-vs-eta}.
With the inclusion of sub-nucleonic fluctuations, the energy carried by gluons increases (due to the increase of $K_g$) and tends to concentrate in the mid-rapidity region while the energy carried by quarks decreases.
When considering the baryon number deposition, we find that the thickness fluctuations suppress baryon stopping slightly while the situations with 3 hotspots suppress baryon stopping significantly, as the thickness profiles of the two colliding nuclei are less likely to overlap.

% \begin{figure}
%     \centering
%     \includegraphics[width=\linewidth]{initial_e2_vs_eta.png}
%     \caption{ The eccentricity of initial gluon and quark along the pseudo-rapidity direction, for Pb+Pb $\sqrt{s_{NN}}$ = 2.76 TeV collisions with centrality range 0-5\%, 5-10\%, 10-15\%, 20-25\%, 30-35\% and 40-45\%.}
%     \label{fig:e2-vs-eta}
% \end{figure}

% \begin{figure}
%     \centering
%     \includegraphics[width=\linewidth]{initial_r2_vs_eta.png}
%     \caption{ The initial longitudinal decorrelations of eccentricity for Pb+Pb $\sqrt{s_{NN}}$ = 2.76 TeV collisions with centrality range 0-5\%, 5-10\%, 10-15\%, 20-25\%, 30-35\% and 40-45\%}
%     \label{fig:r2-vs-eta}
% \end{figure}

\subsection{(3+1)D viscous hydrodynamics}
\label{hydro}
Initial energy(or entropy) deposition models are typically connected to a viscous hydrodynamic model to describe the dynamical evolution of heavy-ion collisions, as in Refs.~\cite{Zhao:2017yhj, Ding:2021ajz, Wu:2018cpc, Chen:2024xbi}.
In this work, the (3+1)-dimensional CLVisc viscous hydrodynamics model \cite{Wu:2021fjf, Pang:2018zzo} is used to simulate the space-time evolution of the QGP fireball. The following energy-momentum and baryon number conservation equations are solved: 
\begin{equation}
\begin{aligned}
    \partial_{\mu} T^{\mu\nu} = 0,\\
    \partial_{\mu} J_B^{\mu} = 0.
\end{aligned}
\end{equation} 
Since the QGP medium is not fully in local thermal equilibrium, we also solve the evolution of shear tensor, bulk pressure and baryon diffusion current using the Israel-Stewart-like equations~\cite{Denicol:2018wdp},
\begin{equation}
\begin{gathered}
\tau_{\Pi} D \Pi+\Pi=-\zeta \theta-\delta_{\Pi \Pi} \Pi \theta+\lambda_{\Pi \pi} \pi^{\mu \nu} \sigma_{\mu \nu} \\
\tau_\pi \Delta_{\alpha \beta}^{\mu \nu} D \pi^{\alpha \beta}+\pi^{\mu \nu}=\eta_v \sigma^{\mu \nu}-\delta_{\pi \pi} \pi^{\mu \nu} \theta-\tau_{\pi \pi} \pi^{\lambda\langle\mu} \sigma_\lambda^{\nu\rangle}\\+\varphi_1 \pi_\alpha^{\langle\mu} \pi^{\nu\rangle \alpha}.\\
\tau_V \Delta^{\mu \nu} D V_\nu+V^\mu=\kappa_B \nabla^\mu \frac{\mu_B}{T}-\delta_{q q} V^\mu \theta-\lambda_{q q} V_\nu \sigma^{\mu \nu}
\end{gathered}
\end{equation}

\begin{table}[t]
\begin{tabular}{|c|c|c|c|}
%\hline
\hline$\tau_{\Pi}$ & $\delta_{\Pi \Pi}$ & $\lambda_{\Pi \pi}$ & \\
\hline$\frac{\zeta(T)}{15\left(\frac{1}{3}-c_s^2\right)^2(e+p)}$ & $\frac{2}{3} \tau_{\Pi}$ & $\frac{4}{5}\left(1-c_s^2\right) \tau_{\Pi}$ & \\
\hline
\hline$\tau_\pi$ & $\delta_{\pi \pi}$ & $\tau_{\pi \pi}$ & $\varphi_1$ \\
\hline$\frac{5C_{\eta_v}}{T}$ & $\frac{4}{3} \tau_\pi$ & $\frac{5}{7}\tau_\pi$ & $\frac{9}{70} \frac{4}{e+p}\tau_\pi$ \\
\hline
\hline$\tau_V$ & $\delta_{qq}$ & $\lambda_{qq}$ & \\
\hline$\frac{C_B}{T}$ & $\tau_V$ & $\frac{3}{10}\tau_V$ & \\
\hline
\end{tabular}
\caption{The relaxation times of shear tensor($\tau_{\pi}$), bulk pressure($\tau_\Pi$) and baryon diffusion current($\tau_V$) as well as respective transport coefficients for the second-order Israel-Stewart equation of motion~\cite{Denicol:2014vaa}.}
\label{tab:relaxation}
\end{table}

Here $D$ is the covariant derivative $u^\mu \partial_\mu$, $\Delta^{\mu\nu}$ is the spatial projector $g^{\mu\nu}-u^\mu u^\nu$, $\nabla^\mu=\Delta^{\mu\nu}\partial_\nu$, $\theta=\partial_{\mu}u^{\mu}$ is the expansion rate and $\sigma^{\mu\nu}$ is the symmetric shear tensor $2\nabla^{<\mu} u^{\nu>}$. 
The traceless tensor $A^{\langle\mu \nu\rangle}=\Delta_{\alpha \beta}^{\mu \nu}A^{\alpha \beta}$ projects out the part that is traceless and transverse to the flow velocity using the double, symmetric, and traceless projection operator $\Delta_{\alpha \beta}^{\mu \nu}=\frac{1}{2}\left(\Delta_\alpha^\mu \Delta_\beta^\nu+\Delta_\alpha^\nu \Delta^\mu{ }_\beta\right)-\frac{1}{3} \Delta^{\mu \nu} \Delta_{\alpha \beta}$. 
$\eta_\nu, \zeta$ and $\kappa_B$ are the transport coefficients for the evolution of shear tensor, bulk pressure and diffusion current. $\tau_{\pi}$, $\tau_\Pi$ and $\tau_V$  are relaxation times of shear tensor, bulk pressure and baryon diffusion current. $\delta_{\Pi \Pi},\lambda_{\Pi \pi},\delta_{\pi \pi},\tau_{\pi \pi},\varphi_1$, $\delta_{qq}$ and $\lambda_{qq}$ are second-order transport coefficients. The values of relaxation time and transport coefficients\cite{Denicol:2014vaa} are summarized in Table \ref{tab:relaxation}.

In this paper, we consider the specific shear viscosity $C_{\eta_\nu}$, bulk viscosity $C_{\zeta}$ and baryon diffusion coefficient $C_B$ as model parameters, which are related to $\eta_\nu$, $\zeta$
and $\kappa_B$ as follows:
\begin{equation}
\begin{aligned}
C_{\eta_v} & =\frac{\eta_v T}{e+P},\quad \qquad C_{\zeta} =\frac{\zeta T}{e+P} \\
\kappa_B & =\frac{C_B}{T} n\left[\frac{1}{3} \coth \left(\frac{\mu_B}{T}\right)-\frac{n T}{e+P}\right] .
\end{aligned}
\end{equation}
The temperature dependence of specific shear viscosity $C_{\eta_v}$ and bulk viscosity  $C_{\zeta}$ are taken from~\cite{Bernhard:2016tnd}, and we set
$C_B=0.4$. The equation of state including baryon potential(NEOS-B)\cite{Monnai:2019hkn} is utilized.

\subsection{Particlization}
\label{freezeout}
When the energy density of QGP fluid cell drops down to the freezeout density ($\epsilon_\mathrm{frz}$=0.266477 GeV/fm$^3$), the momentum distribution of hadrons is obtained using the Cooper-Frye prescription,
\begin{equation}
\frac{d N_i}{d y p_T d p_T d \phi}=\frac{g_i}{(2 \pi)^3} \int_\Sigma p^\mu d \Sigma_\mu f_{\mathrm{eq}}(1+\delta f_\pi + \delta f_V),
\end{equation}
where the thermal equilibrium term $f_{\rm {eq}}$ and out-of-equilibrium correction $\delta f$ are given as,
\begin{equation}
\begin{aligned}
f_{\mathrm{eq}} & =\frac{1}{\exp \left[\left(p\cdot u - B \mu_B\right) / T\right] \pm 1}, \\
\delta f_\pi & =(1 \pm f^{\mathrm{eq}}) \frac{p_\mu p_\nu \pi^{\mu \nu}}{2 T^2(e+P)}, \\
\delta f_V & =(1 \pm f^{\mathrm{eq}})\left(\frac{n}{e+P}-\frac{B}{p\cdot u}\right) \frac{p^\mu V_\mu}{\kappa_B / \tau_V},
\end{aligned}
\end{equation}
where B is the baryon number for the identified baryon.
In this work, we neglect the correction from bulk pressure due to its theoretical uncertainty\cite{JETSCAPE:2020mzn,JETSCAPE:2020shq}. 
After the thermal hadrons are produced, they are forced to decay into stable particles immediately instead of employing a hadronic afterburner. 
In order to improve the statistics, we oversample 3000 times for each hydrodynamic event.

\section{Results and Discussions}
\label{results}
We will now present our simulation results for the multiplicity, $p_T$ spectra, anisotropic flow and longitudinal decorrelations in Pb-Pb collisions at $\sqrt{s_{NN}}$=2.76 TeV using the (3+1) dimensional CLVisc hydrodynamics framework with \Dipper\ initial conditions.

\subsection{Multiplicity of charged hadrons and net-proton}
\begin{figure}[t]
    \centering
    \includegraphics[width=0.95\linewidth]{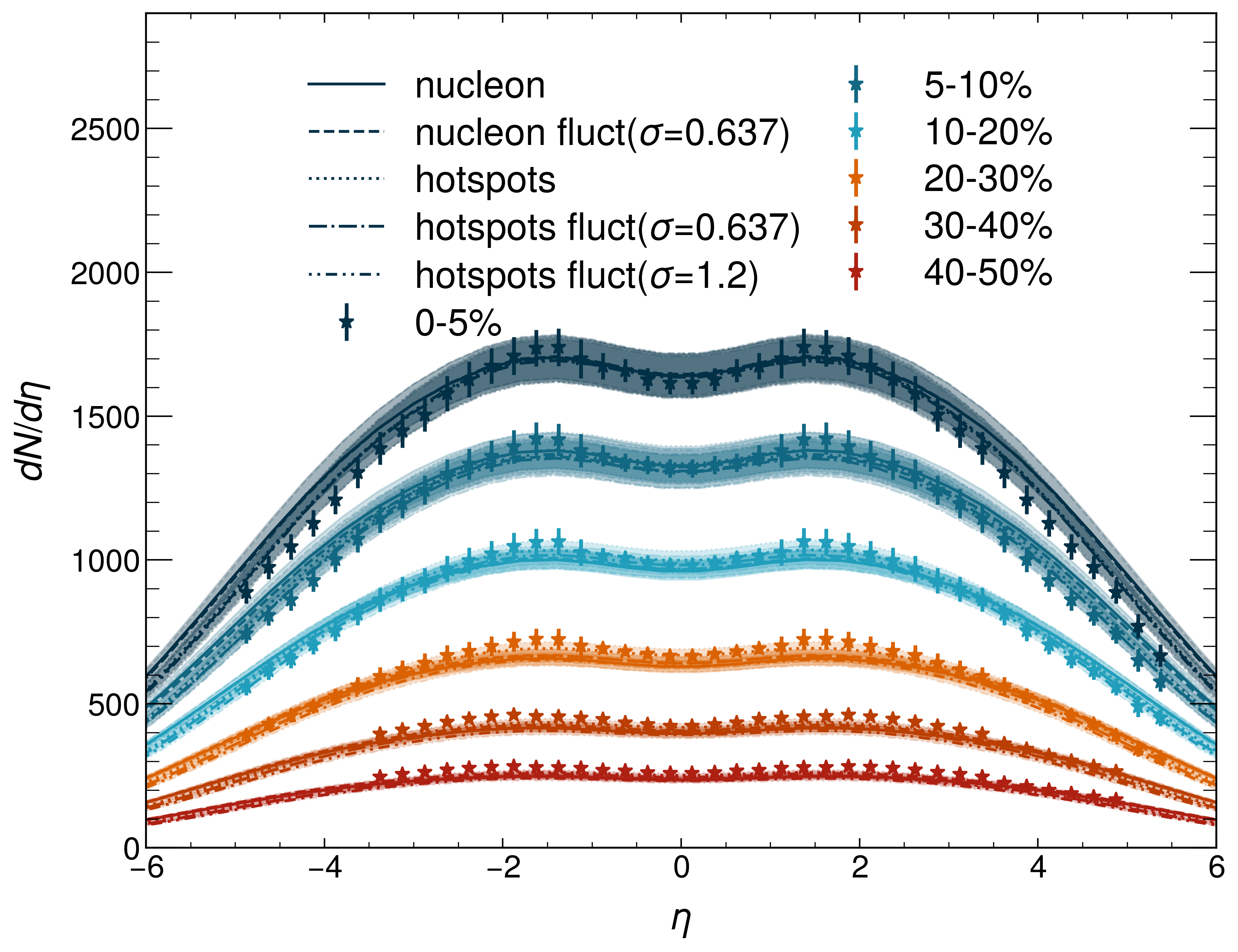}
    \caption{Charged hadron multiplicity as a function of pseudo-rapidity $\eta$ for Pb+Pb collisions at 2.76 TeV from event-by-event hydrodynamics simulation(lines) compared to the ALICE results\cite{ALICE:2015bpk,ALICE:2013jfw}.
    The shaded areas around each line represent the statistical uncertainty of the simulation results.}
    \label{fig:multi-char}
\end{figure}

We first show the charged hadrons pseudo-rapidity distribution in Fig. \ref{fig:multi-char}. Because  in each setup of the \Dipper~the value of $K_g$ is tuned to reproduce the pseudo-rapidity distribution of charged hadrons in central collisions at mid-rapidity, we find that for the most central collisions $dN/d\eta$ of the five different setups is completely identical around mid-rapidity. In the forward or backward region and in other centrality classes, our model can qualitatively describe the ALICE data. We also find that sub-nucleonic fluctuations only have limited effects on the multiplicity distribution, but the inclusion of 3 hotspots inside each nucleon can give a slightly better description of particle production in forward or backward region.

\begin{figure}
    \centering
    \includegraphics[width=0.95\linewidth]{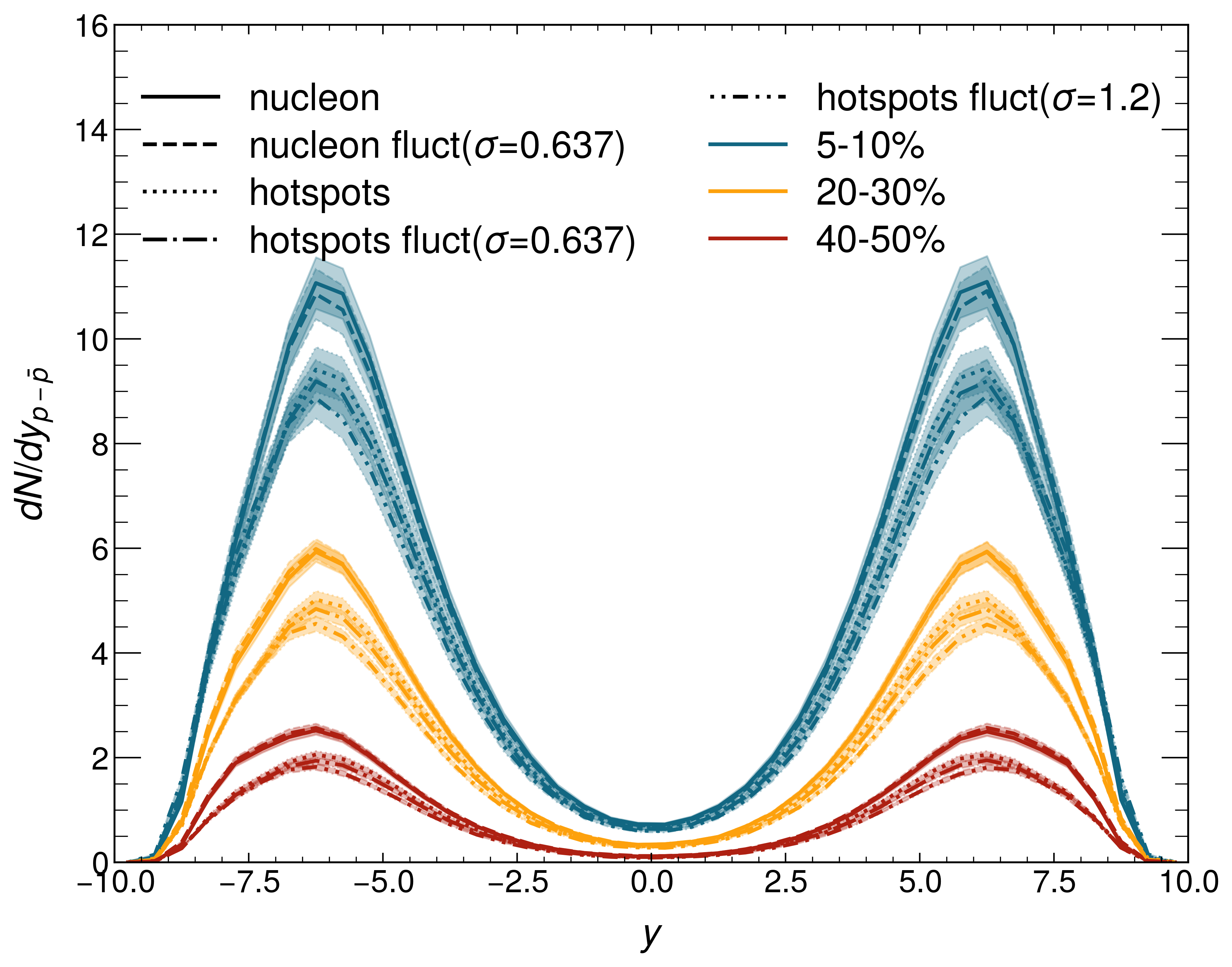}
    \caption{Net proton multiplicity as a function of rapidity $y$ for Pb+Pb collisions at 2.76 TeV from hydrodynamics simulations.
    The shaded areas around each line represent the statistical uncertainty.}
    \label{fig:net-proton}
\end{figure}

In Fig. \ref{fig:net-proton}, we show the net proton multiplicity distribution in rapidity direction in some representative centrality classes. As expected, we obtain less baryon stopping when including more fluctuations which makes sense because there are more empty spots in the thickness profiles of both incoming nuclei.
Overall, in 5-10\% centrality, around 22\% of the baryons are stopped in the nucleon case while only about 19\% of the baryons are stopped in the hotspot case with thickness fluctuations.

\begin{figure}[hbtp]
    \centering
    \includegraphics[width=0.9\linewidth]{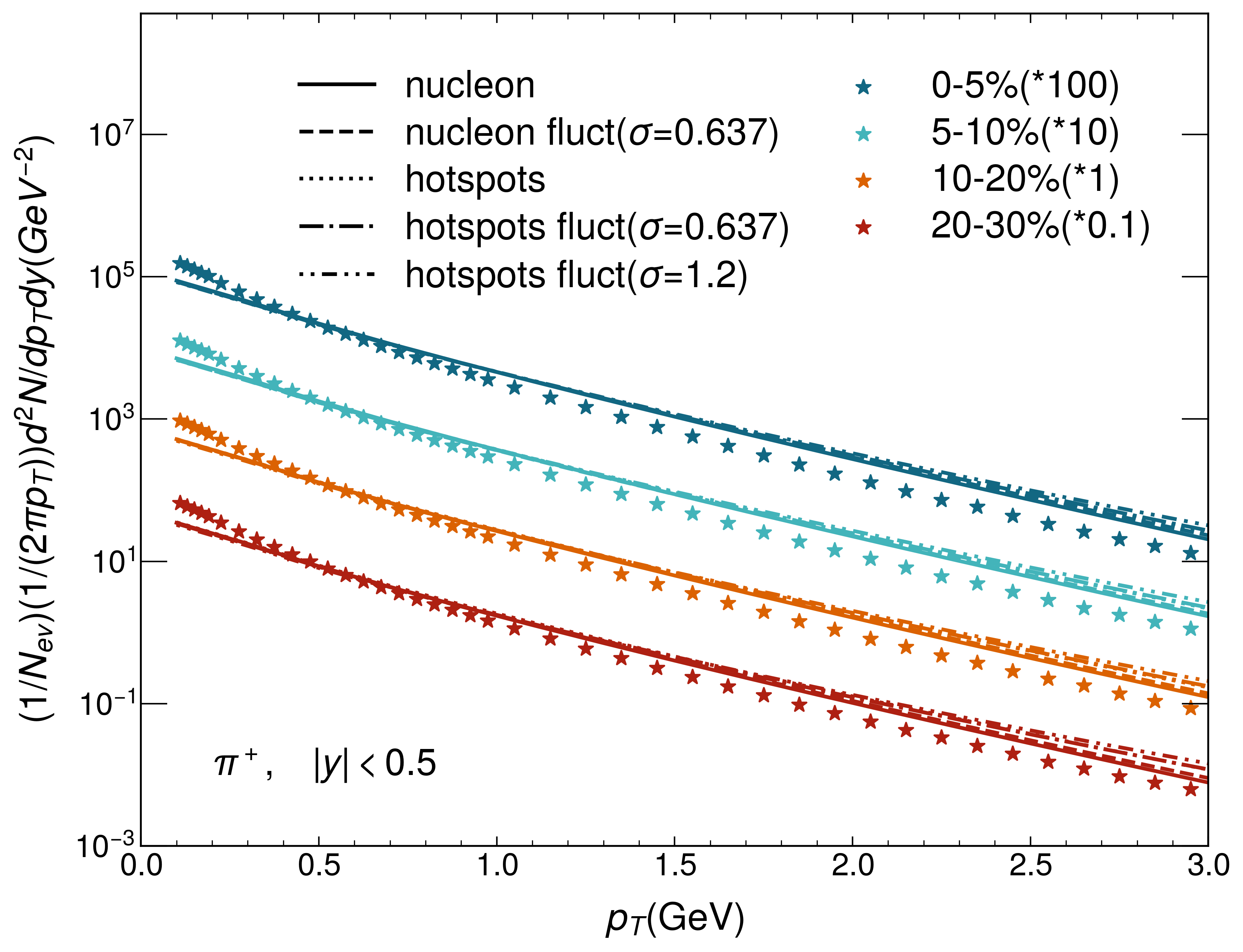}
    \includegraphics[width=0.9\linewidth]{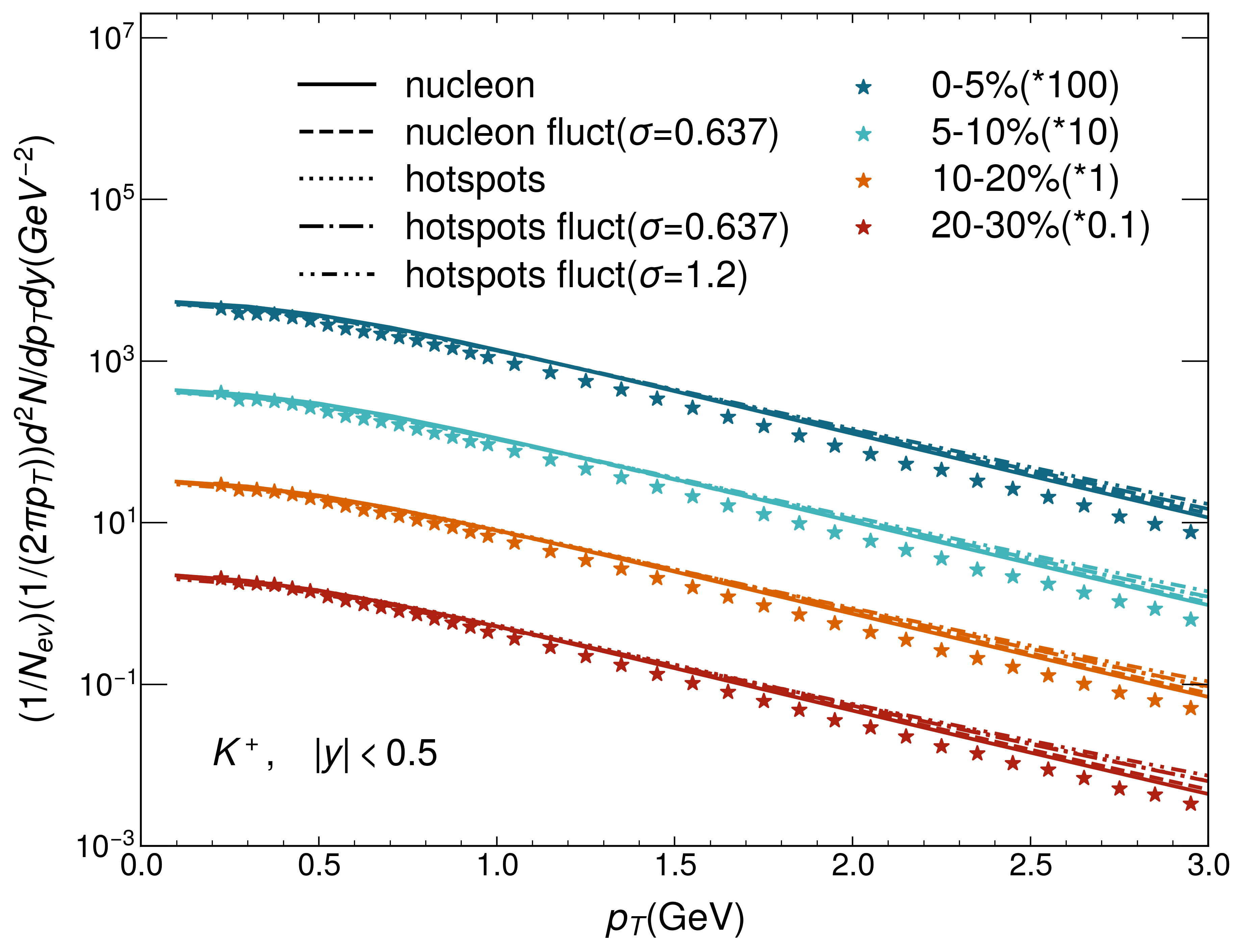}
    \includegraphics[width=0.9\linewidth]{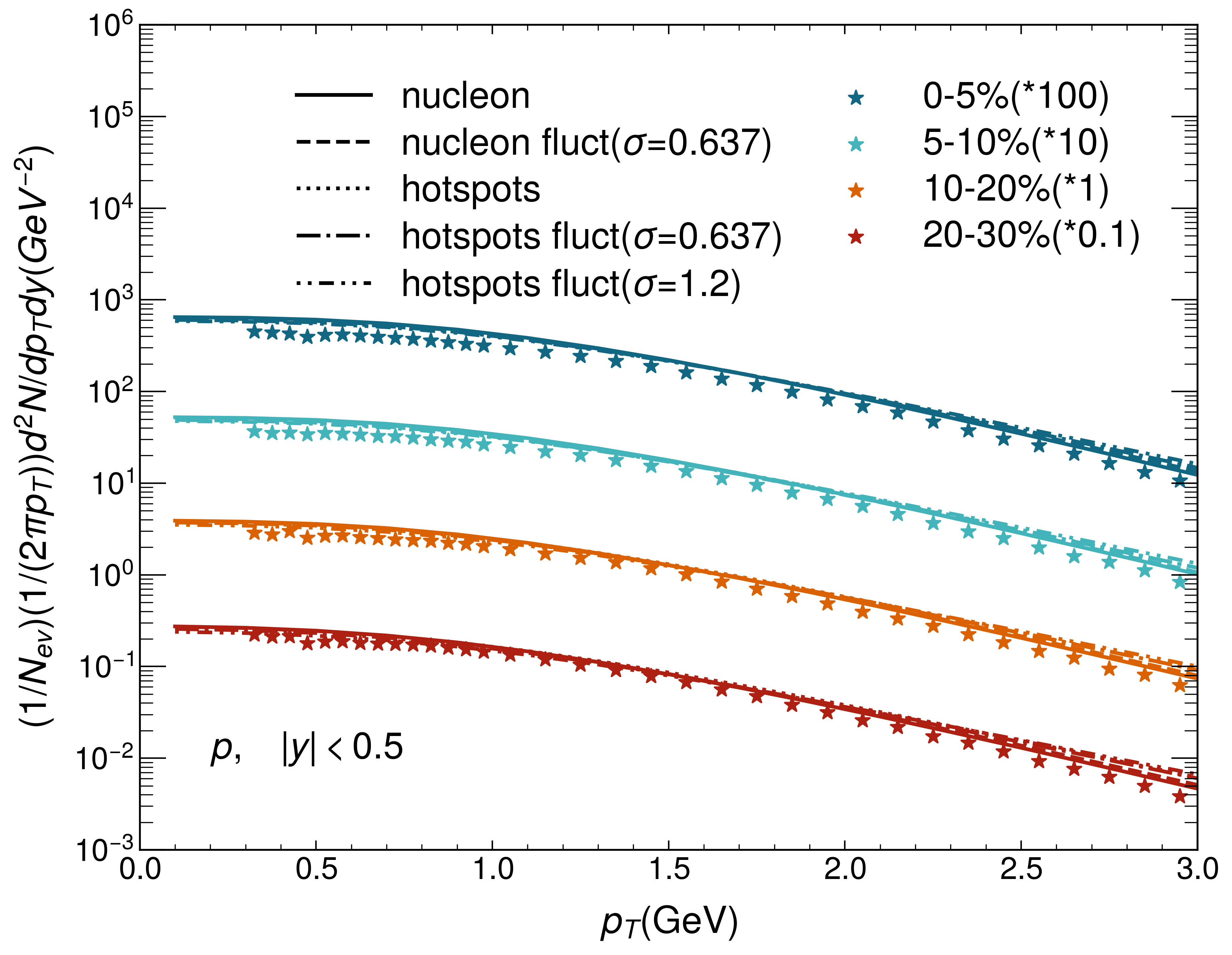}
    \caption{The transverse momentum spectra for identified particles ($\pi^+, K^+, p$) in different centrality classes for Pb+Pb collisions at $\sqrt{s_{NN}}$=2.76 TeV from hydrodynamics simulation(lines). The data(points) are taken from ALICE collaboration\cite{ALICE:2013mez}.}
    \label{fig:ptspec}
\end{figure}

\subsection{Transverse momentum spectra of identified hadrons}
Our results for the transverse momentum spectra of pions $\pi^+$, kaons $K^+$ and protons $p$ are shown in Fig. \ref{fig:ptspec}. 
We find that our model can describe the ALICE data qualitatively, especially for heavier particles. We further notice that the inclusion of sub-nucleonic fluctuations leads to a hardening of the $p_T$ spectra, confirming the trend observed in previous studies, see Refs.~\cite{Noronha-Hostler:2015coa, Andrade:2008xh}. 
While the spectra of pions is comparatively harder than the experimental results, this general trend has also been seen in previous studies~\cite{Schenke:2010nt, Nonaka:2006yn, Song:2010aq}.  
We would like to emphasize that no tuning was performed outside of the initial normalization factor $K_g$, and we believe that an improved description of the experimental data can be achieved with an additional tuning of transport coefficients. 
Moreover, we expect the description of $p_T$ spectra to be improved by a full hybrid simulation with a hadronic afterburner~\cite{Nonaka:2006yn,Song:2010aq}. 

\subsection{Directed flow}
\begin{figure*}
    \centering
    \includegraphics[width=0.8\linewidth]{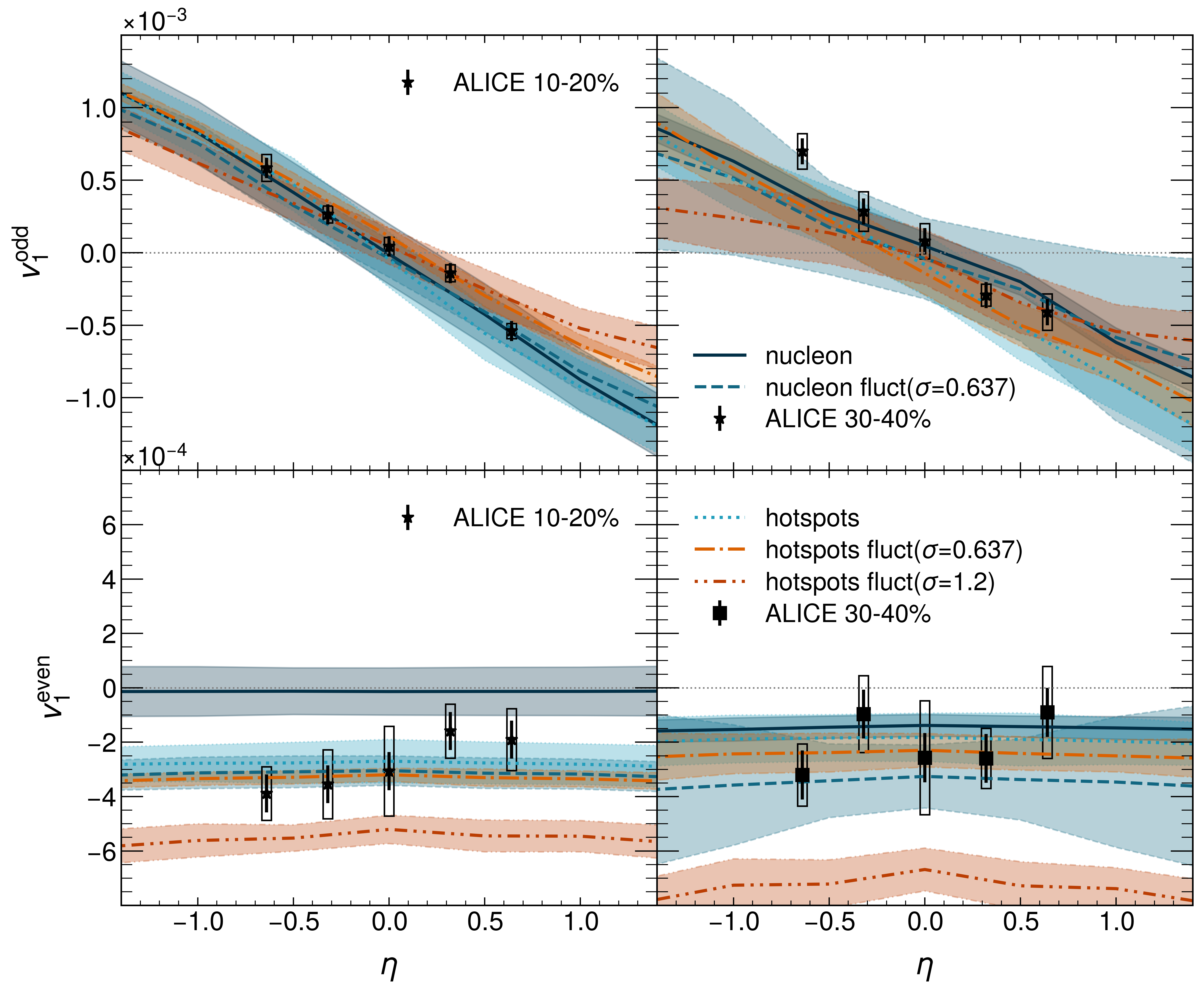}
    \caption{Directed flow $v_1(\eta)$ of charged particles from \Dipper(curves)+CLVisc for Pb+Pb $\sqrt{s_{NN}}$=2760 GeV collisions in comparison to the experiment data from ALICE Collaboration(solid points)\cite{ALICE:2013xri}.
    The shaded areas around each line represent the statistical uncertainty of the simulation results.}
    \label{fig:v1}
\end{figure*}
Next we continue to analyze the development of anisotropic flow, as quantified by the anisotropic flow coefficients $v_n$. The first coefficient of anisotropic flow, $v_1$, or directed flow, as it is usually called in the literature,  reflects the response to the "tilt" of the fireball along the rapidity direction~\cite{Bozek:2010bi,Bozek:2011ua}. In this work, the directed flow is calculated relative to the spectator plane as in \cite{ALICE:2013xri},
\begin{equation}
\begin{aligned}
v_1\left\{\Psi_{\mathrm{SP}}^{\mathrm{p}}\right\}=\frac{1}{\sqrt{2}}
\left[\frac{\left\langle u_x \mathrm{Q}_x^{\mathrm{p}}\right\rangle}{\sqrt{\left|\left\langle\mathrm{Q}_x^{\mathrm{t}} \mathrm{Q}_x^{\mathrm{p}}\right\rangle\right|}}+\frac{\left\langle u_y \mathrm{Q}_y^{\mathrm{p}}\right\rangle}{\sqrt{\left|\left\langle\mathrm{Q}_y^{\mathrm{t}} \mathrm{Q}_y^{\mathrm{p}}\right\rangle\right|}}\right],\\
v_1\left\{\Psi_{\mathrm{SP}}^{\mathrm{t}}\right\}=-\frac{1}{\sqrt{2}}
\left[\frac{\left\langle u_x \mathrm{Q}_x^{\mathrm{t}}\right\rangle}{\sqrt{\left|\left\langle\mathrm{Q}_x^{\mathrm{t}} \mathrm{Q}_x^{\mathrm{p}}\right\rangle\right|}}+\frac{\left\langle u_y \mathrm{Q}_y^{\mathrm{t}}\right\rangle}{\sqrt{\left|\left\langle\mathrm{Q}_y^{\mathrm{t}} \mathrm{Q}_y^{\mathrm{p}}\right\rangle\right|}}\right],
\end{aligned}
\end{equation}
where $u_x=\cos\phi$ and $u_y=\sin\phi$ are defined for charged particles located in the rapidity region of interest,
the brackets "$\langle...\rangle$" indicate an average over measured particles in all recorded events and p/t denotes projectile(target).
Spectator deflection vectors $\mathbf{Q}^{\mathrm{t}, \mathrm{p}}$ are estimated from the positions of spectator nucleons from \Dipper\ as
$$
\mathbf{Q}^{\mathrm{t}, \mathrm{p}} 
\equiv\left(\mathrm{Q}_x^{\mathrm{t}, \mathrm{p}}, \mathrm{Q}_y^{\mathrm{t}, \mathrm{p}}\right)
=\frac{1}{N^{t,p}_{\rm spec}}\sum (\mathbf{r}_{\perp,i}-\mathbf{r}_{\rm com}),
$$
where $r_{\perp,i}$ are the transverse position of spectator nucleons, $r_{\rm com}$ is the center of mass of initial fireball.
Finally, we take the average or difference between $v_1\{\Psi_{\rm SP}^p\}$ and $v_1\{\Psi_{\rm SP}^t\}$ to get $v_1^{\rm odd}\{\Psi_{\rm SP}\}$ and $v_1^{\rm even}\{\Psi_{\rm SP}\}$
\begin{equation}
\begin{aligned}
&v_1^{\rm odd}\{\Psi_{\rm SP}\}
=\left[v_1\left\{\Psi_{\mathrm{SP}}^{\mathrm{p}}\right\}+v_1\left\{\Psi_{\mathrm{SP}}^{\mathrm{t}}\right\}\right]/2,\\
&v_1^{\rm even}\{\Psi_{\rm SP}\}
=\left[v_1\left\{\Psi_{\mathrm{SP}}^{\mathrm{p}}\right\}-v_1\left\{\Psi_{\mathrm{SP}}^{\mathrm{t}}\right\}\right]/2,
\end{aligned}
\end{equation}

By looking at the results depicted in Fig. \ref{fig:v1}, we find that our model can describe the ALICE data regarding the odd components of directed flow really well while the even components are underestimated when the sub-nucleonic fluctuations are absent. This is because the $v_1^{\rm odd}$ mainly comes from the collision geometry in reaction plane~\cite{Bozek:2010bi}, however, the $v_1^{\rm even}$ originates from dipole-like event-by-event fluctuations in the initial state~\cite{Luzum:2010fb} which should be sensitive to the granularity of fluctuations.

With regards to the effects of sub-nucleonic fluctuations, we find that simulations with 3 hotspots inside each nucleon tend to enhance $v_1^{\rm odd}$ while the thickness fluctuation is inclined to suppress $v_1^{\rm odd}$. Nevertheless, both hotspots and thickness fluctuations tend to enhance $v_1^{\rm even}$.

\subsection{Elliptic and triangular flow}\label{subsec-v2}
\begin{figure*}
    \centering
    \includegraphics[width=0.9\textwidth]{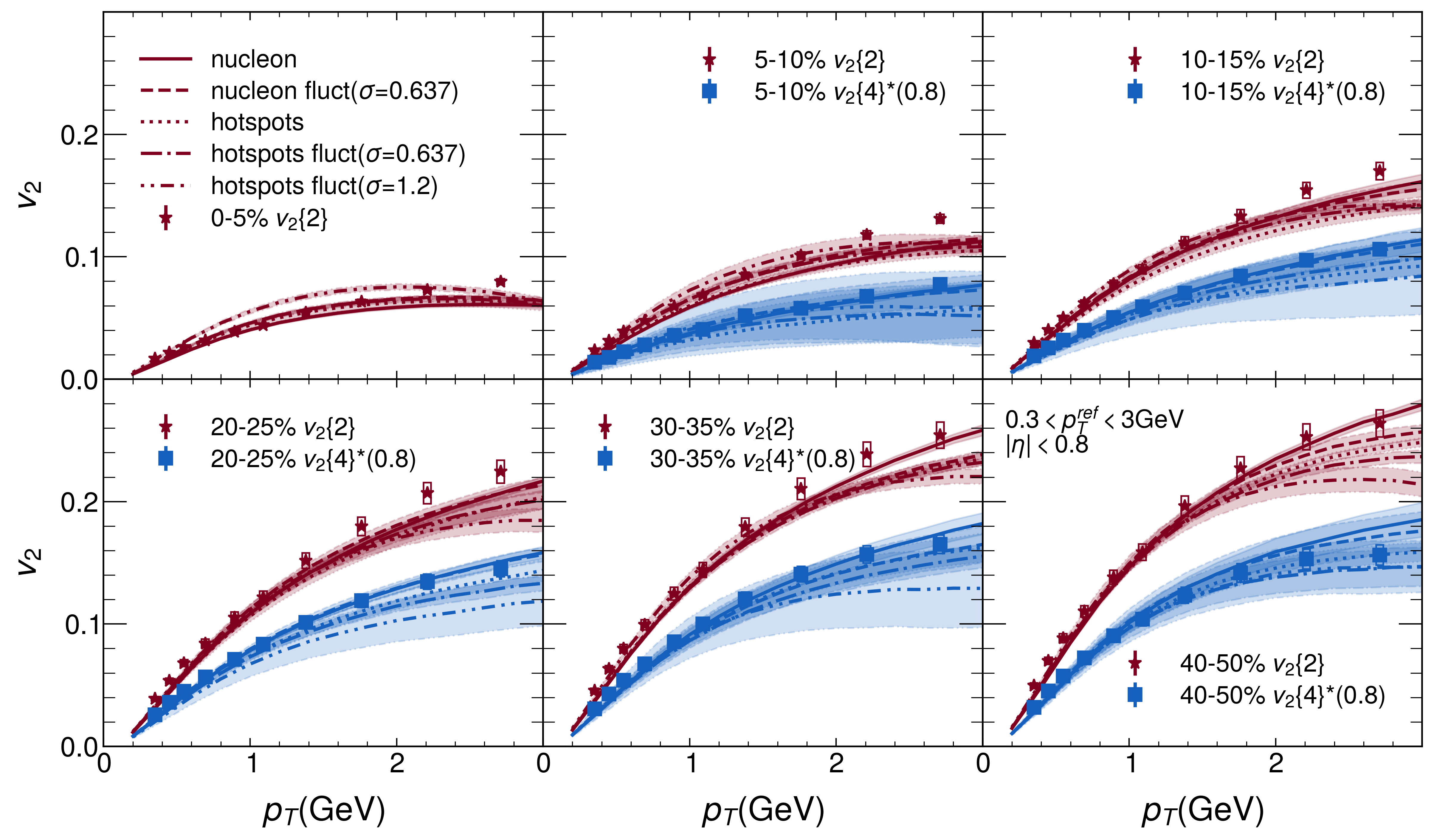}
    \caption{The elliptic flow, $v_2\{2\}$(red curves) and ${v_{{2}}\{4\}}$(blue curves), for charged particles  as a function of transverse momentum $p_T$  in Pb+Pb collisions at $\sqrt{s_{NN}}$ = 2.76 TeV. The data(points) are taken from CMS~\cite{CMS:2012zex}. 
    The shaded areas around each line represent the statistical uncertainty of the simulation results.}
    \label{fig:v2pt-cum}
\end{figure*}

Elliptic flow is the most widely studied observable in heavy-ion collisions which is usually analyzed with the so-called multi-particle cumulant method \cite{Borghini:2001vi,Borghini:2000sa,Bilandzic:2010jr}. The differential flow $v_n^\prime\{2\}$ and $v_n^\prime\{4\}$ calculated with two- and four-particle cumulants are defined as
\begin{equation}
\begin{split}
v_n^{\prime}\{2\}
&=\frac{d_n\{2\}}{\sqrt{c_n\{2\}}}
=\frac{\langle\!\langle 2^\prime \rangle\!\rangle}
{\sqrt{\langle\!\langle 2 \rangle\!\rangle}}\,,\\
v_n^{\prime}\{4\}&=-\frac{d_n\{4\}}{\left(-c_n\{4\}\right)^{3 / 4}}
    =-\frac{\langle\!\langle 4^{\prime}\rangle\!\rangle-2 \cdot\langle\!\langle 2^{\prime}\rangle\!\rangle\langle\!\langle 2\rangle\!\rangle}{\left( 2 \cdot\langle\!\langle 2\rangle\!\rangle^2-\langle\!\langle 4\rangle\!\rangle \right)^{3 / 4}},
\end{split}
\label{eq:vn_pc}
\end{equation}

where the 2(4)-particle differential cumulants $d_n\{2\}$ and $d_n\{4\}$~\cite{Borghini:2001vi, Bilandzic:2010jr} are defined by correlators
\begin{equation}
\left\langle 2^{\prime}\right\rangle \equiv\left\langle e^{i n\left(\psi_1-\phi_2\right)}\right\rangle  
, \quad
   \left\langle 4^{\prime}\right\rangle \equiv\left\langle e^{i n\left(\psi_1+\phi_2-\phi_3-\phi_4\right)}\right\rangle,
\end{equation}
while the 2(4)-particle integrated cumulants $c_n\{2\}$ and $c_n\{4\}$~\cite{Borghini:2001vi, Bilandzic:2010jr} are defined by correlators
\begin{equation}
  \langle 2\rangle \equiv\left\langle e^{i n\left(\phi_1-\phi_2\right)}\right\rangle
  ,\quad
    \langle 4\rangle \equiv\left\langle e^{i n\left(\phi_1+\phi_2-\phi_3-\phi_4\right)}\right\rangle.
\end{equation}
Here, $\phi$ is azimuthal angle of reference particle(covering a broad rapidity range) while $\psi$ is azimuthal angle of the particle of interest(located in a narrow rapidity range). $\langle ...\rangle$ means an average over particles in a event while $\langle\langle ...\rangle\rangle$ denotes an average, first over all particles and then over all events.
\begin{figure*}
    \centering
    \includegraphics[width=0.9\textwidth]{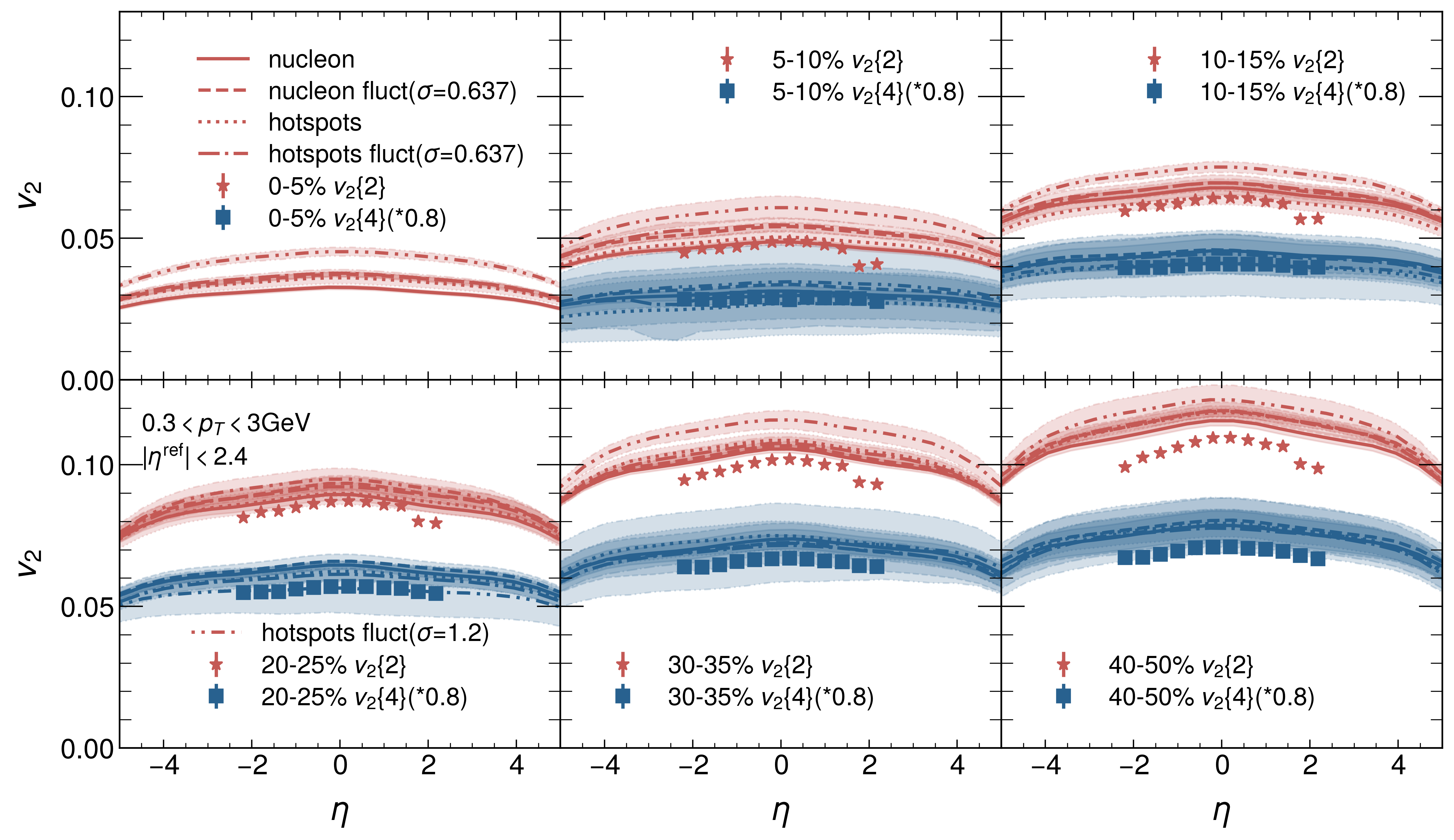}
    \caption{The rapidity dependence of the momentum anisotropies $v_2(\eta)$, compared to CMS\cite{CMS:2012zex} data. 
    Both data and the calculation are for 0.3<$p_T$<3 GeV and use reference particles in (|$\eta$|<2.4).
    The shaded areas around each line represent the statistical uncertainty of the simulation results.}
    \label{fig:v2eta-cms}
\end{figure*}

\begin{figure*}
    \centering
    \includegraphics[width=0.9\textwidth]{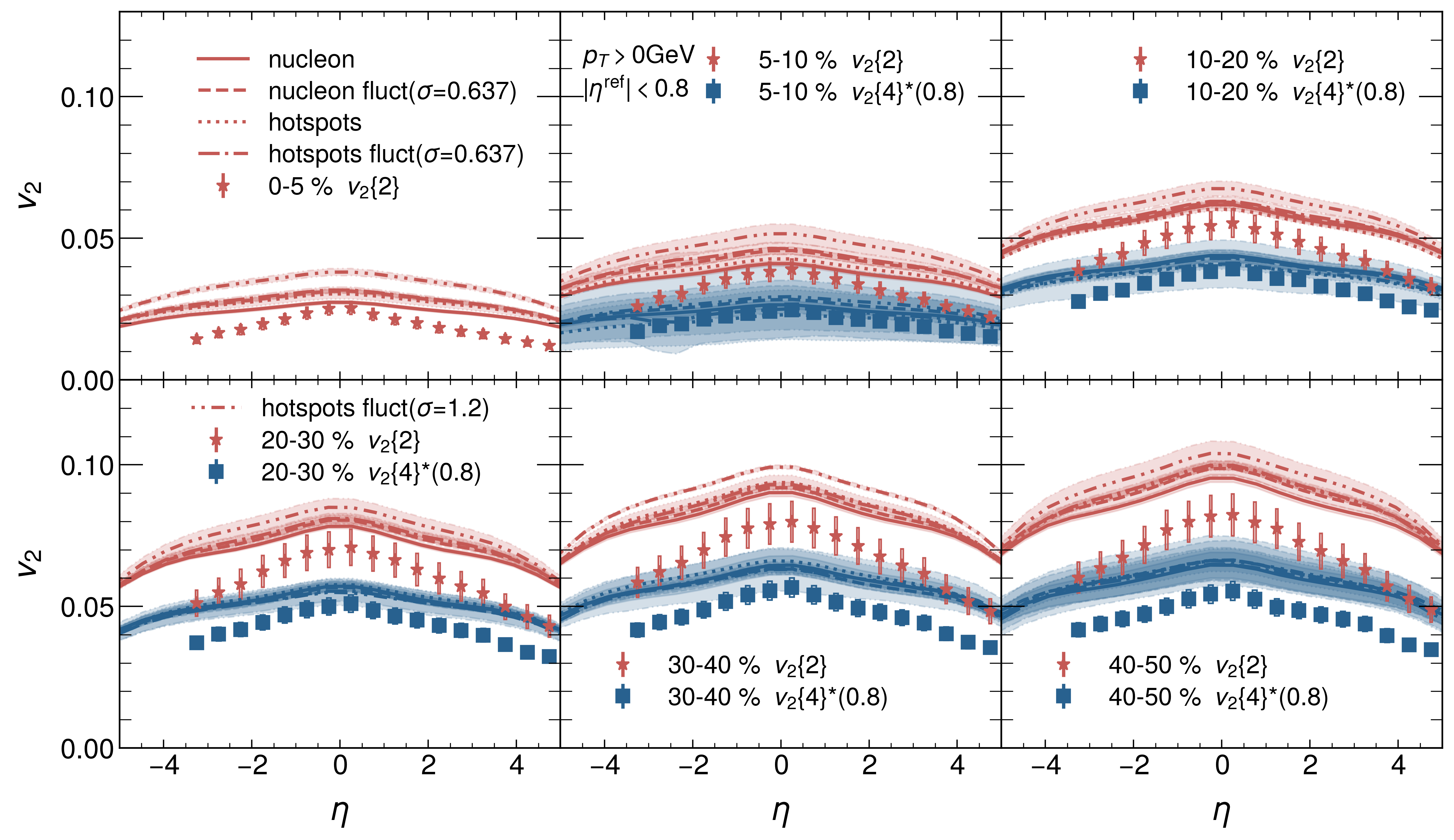}
    \caption{The rapidity dependence of the momentum anisotropies $v_2(\eta)$, compared to ALICE\cite{ALICE:2016tlx} data. Both data and the calculation are for $p_T$>0 GeV and use reference particles at mid-rapidity (|$\eta$|<0.8).
    The shaded areas around each line represent the statistical uncertainty of the simulation results.}
    \label{fig:v2eta-alice}
\end{figure*}

In Fig. \ref{fig:v2pt-cum}, we present the $p_T$-dependence of elliptic flow at mid-rapidity in Pb-Pb collisions at $\sqrt{s_{NN}}$=2.76 TeV. We can find that the integrated \Dipper+CLVisc model generally works well at low $p_T$ to describe the experimental data. As more fluctuations are included in the initial conditions, the flows in central collisions are enhanced while the flows in mid-central or peripheral collisions are suppressed, as additional sources of fluctuations add perturbations to the overall geometry. We have also evaluated the $p_T$-dependence of elliptic flow as computed with the event-plane method, as shown in Fig. \ref{fig:v2pt-ep} of Appendix \ref{app-v2}) and observe the same overall trend as the one presented here with the particle cumulant method.

Next, we move on to investigate the longitudinal structure of flow harmonics in Pb-Pb collisions, and first compare the $\eta$-dependence of elliptic flow with experimental results from CMS and ALICE collaborations in Figs.~\ref{fig:v2eta-cms} and \ref{fig:v2eta-alice}. 
Overall, our simulations exhibit similar shapes as experimental results but overestimate the magnitude, which may be attributed to the too hard $p_T$ spectra in our calculation.

%{\color{red} On the other hand, it is also possible that the overestimation of $v_2(\eta)$ arises from the underestimation of decorrelation between the particles in forward(backward) and central regions.}

The most important difference between results from CMS (Fig.\ref{fig:v2eta-cms}) and ALICE (Fig.\ref{fig:v2eta-alice}) is the kinetic range of reference particles. For the ALICE data, the reference particles are located in a more central region($|\eta|<0.8$) as compared to the $|\eta|<2.4$ region for CMS. Due to the longitudinal decorrelations embedded in the definition of the multi-particle correlation method, this has the consequence that the flow harmonics decrease faster when moving towards the forward or backward rapidity regions, as previously observed with IP-Glasma and MUSIC \cite{McDonald:2023qwc} which provide similar results as our \Dipper\ and CLVisc model. However, in the context of our analysis, it is important to point out that the longitudinal decorrelation is further complicated by the fact that energy and momentum are carried initially by quarks and gluons (see Fig. \ref{fig:ed-vs-eta}), such that for increasingly large rapidity separations, the longitudinal (de-)correlations become increasingly sensitive to the fluctuations in the quark sector, via the energy deposited due to quark stopping -- as opposed to only gluon radiation.

When sub-nucleonic fluctuations are included, we observe an increase in the overall normalization of the elliptic flow. On the other hand, the inclusion of the hotspots decreases the slope of $v_2(\eta)$, bringing the shape of the rapidity dependence of elliptic flow closer to experimental data. 
In the Fig. \ref{fig:v2eta-cms-ep} of Appendix \ref{app-v2}, we also show the $\eta$-dependence of elliptic flow calculated with the event-plane method which demonstrates similar trends as described here.

\begin{figure*}
    \centering
    \includegraphics[width=0.9\textwidth]{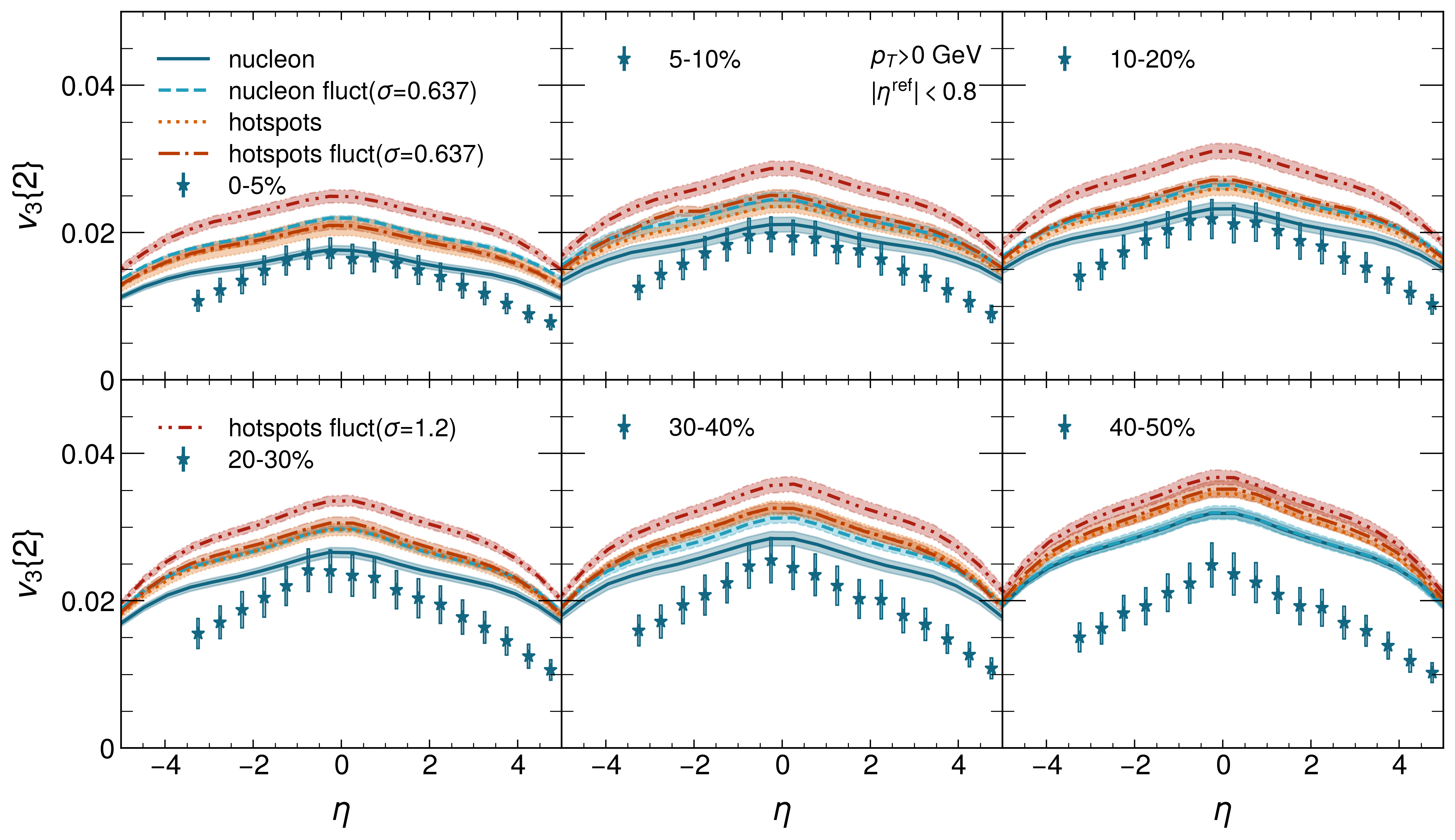}
    \caption{The rapidity dependence of the momentum anisotropies $v_3(\eta)$, compared to ALICE\cite{ALICE:2016tlx} data. Both data and the calculation are for $p_T$>0 GeV and use reference particles at mid-rapidity (|$\eta$|<0.8).
    The shaded areas around each line represent the statistical uncertainty of the simulation results.}
    \label{fig:v3eta-alice}
\end{figure*}

While the elliptic flow $v_2$ receives contributions from the overall event collision geometry, as well as from fluctuations, the higher order odd flow coefficients (e.g. $v_3$) are solely induced by fluctuations. We take $v_3$ as an example to investigate the effect of sub-nucleonic fluctuations on the higher order flow coefficients, and present the corresponding results in Fig. \ref{fig:v3eta-alice}. Similar to the results for elliptic flow, the magnitude of $v_3$ is larger than the experimental results, which again we attribute to the too hard $p_T$ spectrum. However, we can clearly see that as more sub-nucleonic fluctuations are included into the \Dipper, the magnitude of $v_3\{2\}$ gradually increases, while the de-correlation across rapidity also becomes stronger, such that upon including fluctuating hot spots the $\eta$-dependence is similar as in the ALICE data.

\subsection{Flow decorrelation}
In order to further quantify the effects of sub-nucleonic fluctuations on the longitudinal structure of heavy-ion collisions, we now move on to study the longitudinal decorrelation of anistropic flow. We follow the standard procedure, where the longitudinal decorrelation is quantified by the observables $r_n$, which are defined as a forward/backward ratio 
\begin{equation}
r_n\left(\eta^a, \eta^b\right) \equiv \frac{V_{n \Delta}\left(-\eta^a, \eta^b\right)}{V_{n \Delta}\left(\eta^a, \eta^b\right)}
=\frac{\langle Q_n(-\eta^a) Q_n^*(\eta^b)\rangle}{\langle Q_n(\eta^a) Q_n^*(\eta^b)\rangle},
\end{equation}
where ${Q}_n(\eta)=q_n(\eta) e^{i n \Phi_n(\eta)}=\frac{1}{N} \sum_{i=1}^N e^{i n \phi_i}$, $\eta_a$ is the rapidity of interest and $\eta_b$ is the reference rapidity. Intuitively, any deviation of $r_n$ from unity means the decorrelation between the slices located at $\eta_a$ and $-\eta_a$; however as we will see below this observable is also sensitive to the reference rapidity $\eta_b$.

\begin{figure*}
    \centering
    \includegraphics[width=0.9\linewidth]{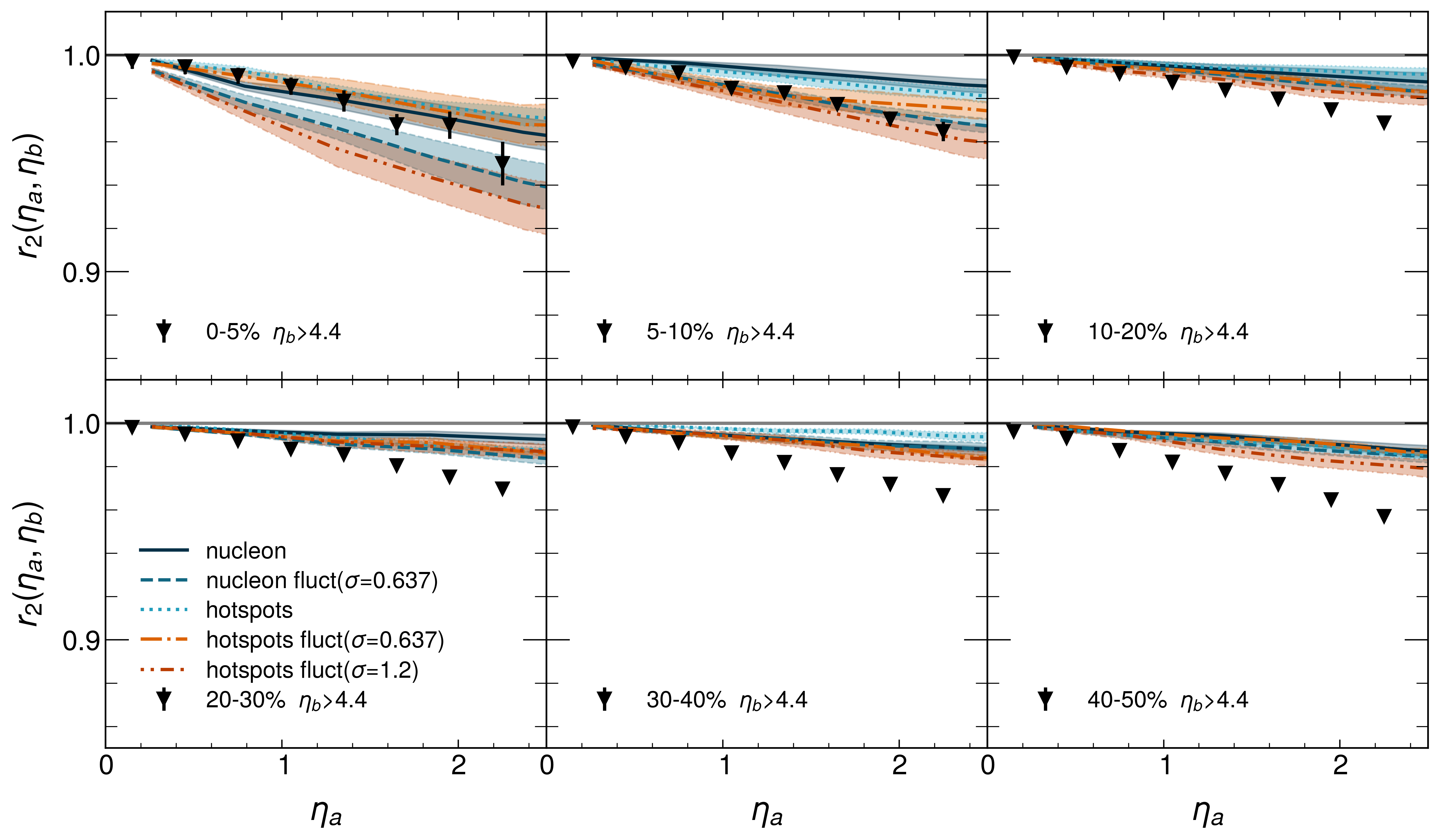}
    \caption{The decorrelation of elliptic flow along the pseudo-rapidity direction, for Pb+Pb $\sqrt{s_{NN}}$ = 2.76 TeV collisions in various centrality classes from (3+1)D viscous hydrodynamic simulations as compared with CMS(black inverted triangles) \cite{CMS:2015xmx}. The reference particles are taken from 4.4<$\eta_b$<5.0.
    The shaded areas around each line represent the statistical uncertainty of the simulation results.}
    \label{fig:r2-b4}
\end{figure*}

\begin{figure*}
    \centering
    \includegraphics[width=0.9\linewidth]{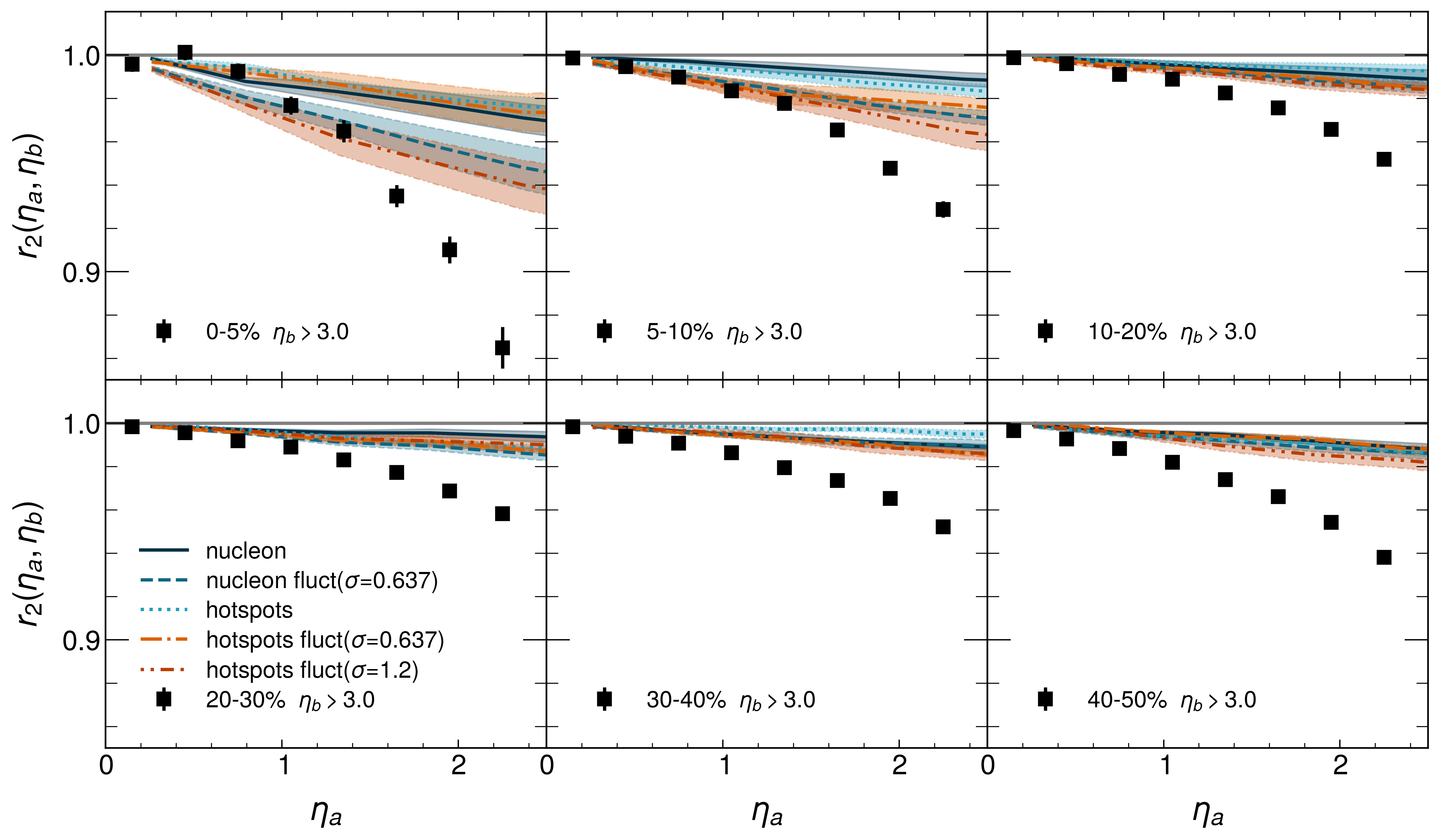}
    \caption{ The decorrelation of elliptic flow along the pseudo-rapidity direction, for Pb+Pb $\sqrt{s_{NN}}$ = 2.76 TeV collisions in various centrality classes from (3+1)D viscous hydrodynamic simulations as compared with CMS(black squares) \cite{CMS:2015xmx}. The reference particles are taken from 3.0<$\eta_b$<4.0.
    The shaded areas around each line represent the statistical uncertainty of the simulation results.}
    \label{fig:r2-b3}
\end{figure*}

We show the decorrelation ratio $r_2(\eta_a)$ of elliptic flow in various centrality classes of Pb-Pb collisions at $\sqrt{s_{NN}}$=2.76 TeV for two different reference ranges $4.4<\eta_b<5$ in Fig. \ref{fig:r2-b4} and $3<\eta_b<4$ in Fig.~\ref{fig:r2-b3}. Before we comment on the model to data comparison, it is interesting to point out, that the experimental results for $3<\eta_b<4$ in Fig.~\ref{fig:r2-b3} show a significantly larger decorrelations compared to the results in Fig. \ref{fig:r2-b4}, which could be partially attributed to the non-flow contamination in experimental measurements. When comparing our \Dipper + CLVisc model to data, we find that for the larger reference rapidity $4.4<\eta_b<5$, the flow decorrelation is well captured for central collisions, and that our model calculations do not exhibit a strong dependence on the reference rapidity $\eta_b$. Importantly, we also find that the inclusion of sub-nucleonic fluctuations in the transverse plane brings about more longitudinal decorrelations, particularly in central collisions. 

When considering mid-central or peripheral events, where elliptic flow is, to a large extent, driven by the geometric eccentricity of the collision, we observe that our model consistently underestimates the longitudinal de-correlation, even upon including sub-nucleonic fluctuations. We believe that this disagreement between simulations and experiments in mid-central (peripheral) collisions should be attributed to missing sources of fluctuations in the initial condition, most likely in the quark sector. Due to the use of collinear parton distributions, the present implementation of the model only considers the energy and charge deposition from an averaged source of quarks.  The suspicion, that missing sources of fluctuations in the quark sector are responsible for underestimating the de-correlation is further corroborated by the fact that the relative
contribution of the energy deposited by quark stopping  to the total energy increases when looking at more peripheral events \cite{Garcia-Montero:2023opu}, and it is also clear that adapting a statistical interpretation of the parton distribution functions as e.g. in Ref.~\cite{Shen:2022oyg}, would lead to further longitudinal decorrelation of the event geometry. 

%We observe that the flow decorrelation is well captured by our {\Dipper}+CLVisc evolution for the central bins and small rapidity separation. In contrast, the decorrelations in mid-central or peripheral events are consistently underestimated. Importantly, we find that the inclusion of sub-nucleonic fluctuations in the transverse plane brings about more longitudinal decorrelations, particularly in central collisions. Nevertheless, even upon including sub-nucleonic fluctuations, our model still underestimates the flow decorrelation in less central collisions, where flow is to a larger extent driven by the overall event geomtry. We believe that this disagreement between simulations and experiments in mid-central (peripheral) collisions should be attributed to missing sources of fluctuations in the initial condition, in particular in quark sector. This is further supported by the fact that the relative contribution of the energy deposited by quark stopping  to the total energy increases when looking at more peripheral events \cite{Garcia-Montero:2023opu}. 

When considering our results in the context of previous studies~\cite{McDonald:2020oyf, Wu:2018cpc}, we observe that initial conditions based on saturation physics, such as IP-Glasma and \Dipper\ usually generate less longitudinal decorrelations than other models, particularly in mid-central and peripheral collisions. However, an important consideration in the context of the \Dipper\ is that -- as already pointed out in Ref.~\cite{Garcia-Montero:2023gex}
-- the physics in forward rapidity region, where energy deposition originates to a significant extent from valence quark stopping, is actually quite different from that around mid-rapidity, where energy deposition is primarily due to gluon radiation. Since the present form of the longitudinal decorrelation measurement, involves both the mid-rapidity region $(\eta_a)$ as well as the far forward region $(\eta_b)$ a thorough description of both regions is required to reproduce the experimental results. Hence in order to further disentangle the interplay of different energy deposition mechanisms, it would also be interesting to 
consider a mid-rapidity reference $\eta_b$, 
albeit this would inevitably require a reasonable treatment to suppress non-flow contributions. 

We close with two remarks. First, a more realistic description of the 3+1D dynamics of the pre-equilibrium stage is desirably to fully understand the evolution of the longitudinal structures imprinted into the QGP. Second, as pointed out in~\cite{Sakai:2021rug, Sakai:2020pjw}, the simultaneous inclusion of initial longitudinal fluctuations and hydrodynamic thermal fluctuations during the evolution of the QGP fluids is essential for a quantitative understanding of the longitudinal decorrelation dynamics.

%{\color{red} STOPPED HERE}

%In the Fig.\ref{fig:r2-b3}, we also show the $r_2(\eta_a)$ with different reference rapidity range $3<\eta_b<4$ where the experimental results show larger decorrelations compared to Fig.\ref{fig:r2-b4} while the simulation results don't change too much. This could be partially attributed to the non-flow contamination in experimental measurements. More importantly, as it was previously emphasized in Ref.~\cite{Garcia-Montero:2023gex}, the physics in forward rapidity is quite different from that in mid-rapidity, so it is important to seek a thorough description of both regions or request for experimental results with different reference kinematics, such as mid-rapidity, with a reasonable treatment of non-flow contribution. 

\section{Summary and outlook}
\label{summary}

In this work, the energy and charge distribution from the {\Dipper} initial state model were, for the first time, connected to the hydrodynamical evolution of the QGP to study the dynamical evolution of HICs at high CoM collisions. This was achieved by interfacing \Dipper\ with the (3+1)-dimensional viscous hydrodynamics model CLVisc. We further extended the initial state model, \Dipper, to include sub-nucleonic fluctuations by considering hotspots and thickness fluctuations for each nucleon . 

With this framework, we analyzed the (pseudo)rapidity distribution of charged particles, baryon stopping, anisotropic flow and longitudinal decorrelation across different centralities in Pb-Pb collisions at $\sqrt{s_{NN}}$=2.76 TeV, and studied the effect of sub-nucleonic fluctuations on longitudinal and transverse observable. Generally, we found that the framework provides rather good descriptions of multiplicity and anisotropic flow in central collisions without any tuning of transport coefficients for hydrodynamics evolution. We further observed that sub-nucleonic fluctuations have evident effects on the $\eta$-dependence of final state observables. More precisely, sub-nucleonic fluctuations can reduce baryon stopping and enhance anisotropic flow and longitudinal decorrelation, especially in central collisions. However,  our framework currently overestimates the magnitude of anisotropic flow and underestimates longitudinal decorrelation for mid-central and peripheral collisions, where anisotropic flow is, to a large extent, geometry driven. We believe that this results likely points to the fact that a further source of fluctuations is missing in the quark sector, and it will be interesting to consider the implementation of valence \cite{Shen:2022oyg} and sea quark fluctuations in the \Dipper\, in the future.

Hence the present work should be understood as a first step towards a more comprehensive ab-initio description of the longitudinal structures in the QGP. While initial state fluctuations inspired by small-$x$ QCD have presently been considered, further work is still needed e.g. to perform a detailed tuning of parameters in order to quantitatively describe the experimental results. Evidently, further theoretical improvements such as the inclusion of valence and sea quark fluctuations, (3+1)D pre-equilibrium dynamics, and the addition of a hadronic afterburner would also be  desirable to provide an accurate description of the longitudinal structure of high-energy heavy-ion collisions. Furthermore, as \Dipper\ has already proven to be an excellent tool for investigating baryon stopping and charge deposition in high energy collisions~\cite{Garcia-Montero:2024jev}, the full event-by-event initial state + pre-equilibrium + hydrodynamic + afterburner simulations will be conducted in the future to thoroughly study the charge evolution of heavy-ion collisions.

\section*{Acknowledgements}
JZ would like to thank Bao-chi Fu, Renata Krupczak, Guang-You Qin, Hendrik Roch, Xiang-Yu Wu for productive discussions. OGM would like to acknowledge productive discussions with Travis Dore, Wilke van der Schee and Govert Nijs. This work is supported by the Deutsche Forschungsgemeinschaft (DFG, German Research Foundation) through the CRC-TR 211 ‘Strong-interaction matter under extreme conditions’ — project number 315477589 — TRR 211. OGM and SS also acknowledge support by the German Bundesministerium für Bildung und Forschung (BMBF) through Grant No. 05P21PBCAA. J.Z. is also supported in part by China Scholarship Council (CSC) under Grant No. 202306770009. 
Numerical simulations presented in this work were performed at the Paderborn Center for Parallel Computing ($\rm PC^2$).

\bibliographystyle{apsrev4-1}

\bibliography{References}

\appendix
\label{appendix}

\section{Initial longitudinal distributions}\label{app-ini}
\begin{figure}[h]
    \centering
    \includegraphics[width=\linewidth]{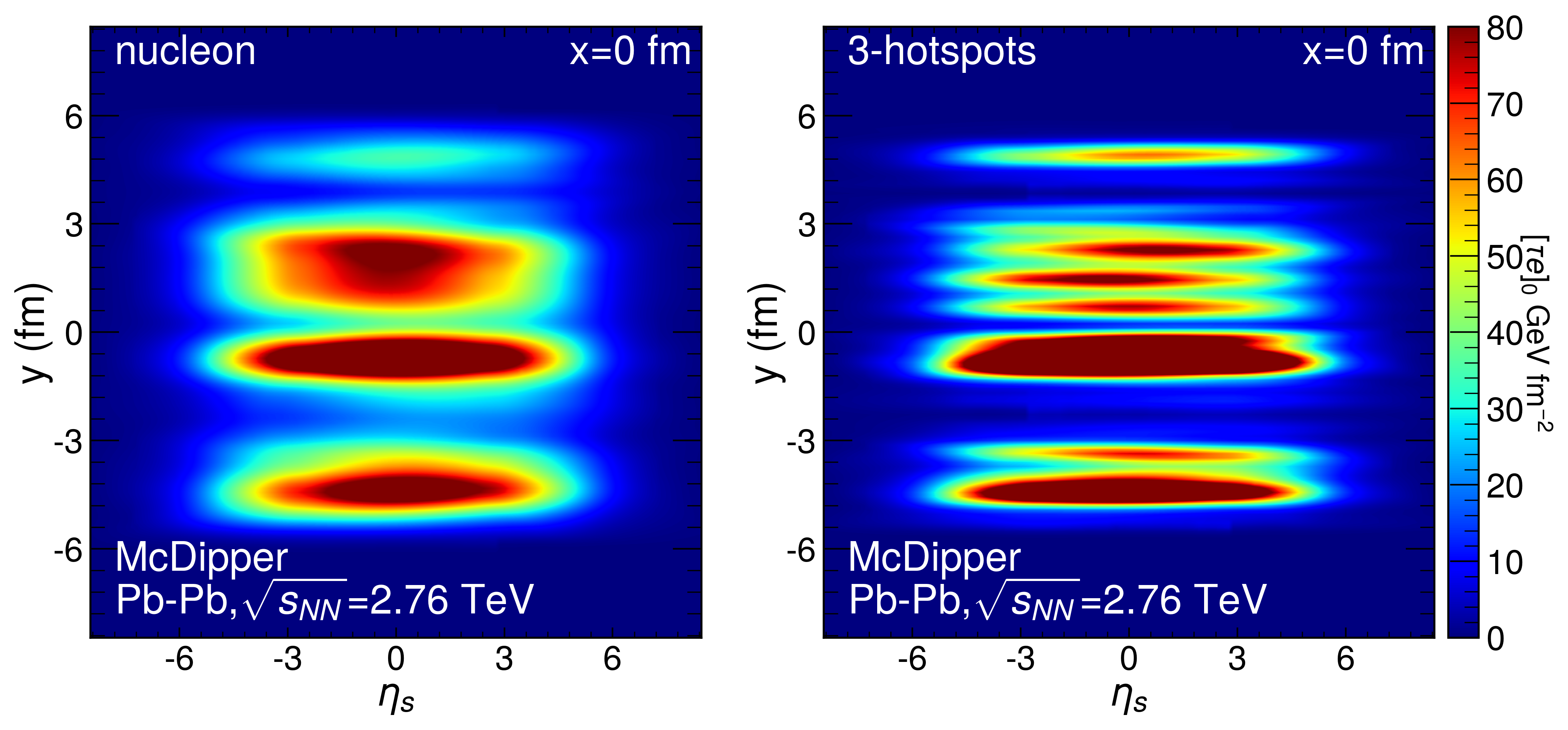}
    \includegraphics[width=\linewidth]{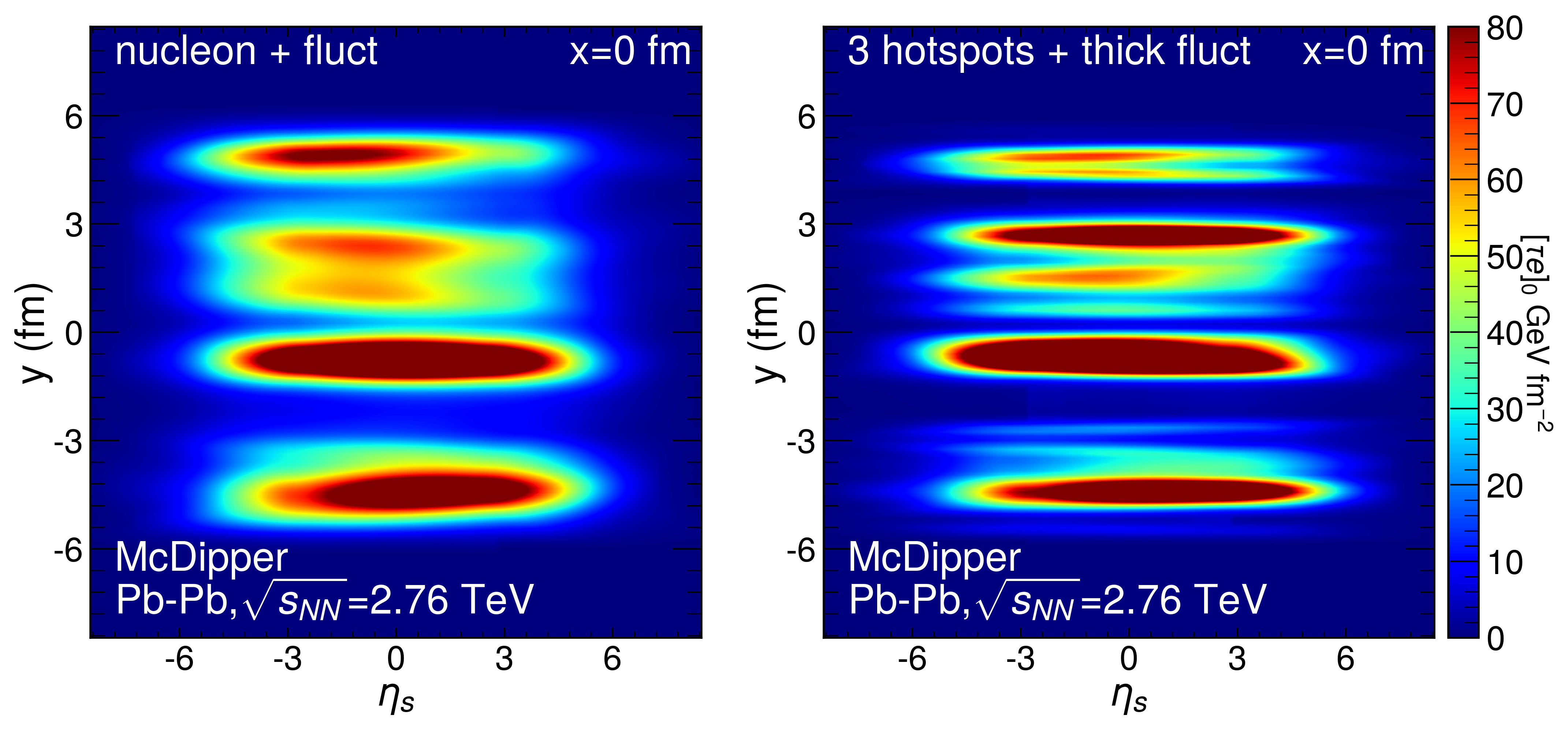}
    \caption{The initial energy density distributions of different classes at x=0 fm for a sampled Pb+Pb collision at $\sqrt{s_{NN}}$=2.76 TeV and impact parameter b=3.42 fm.}
    \label{fig:ini-ed-long}
\end{figure}

\begin{figure}[h]
    \centering
    \includegraphics[width=\linewidth]{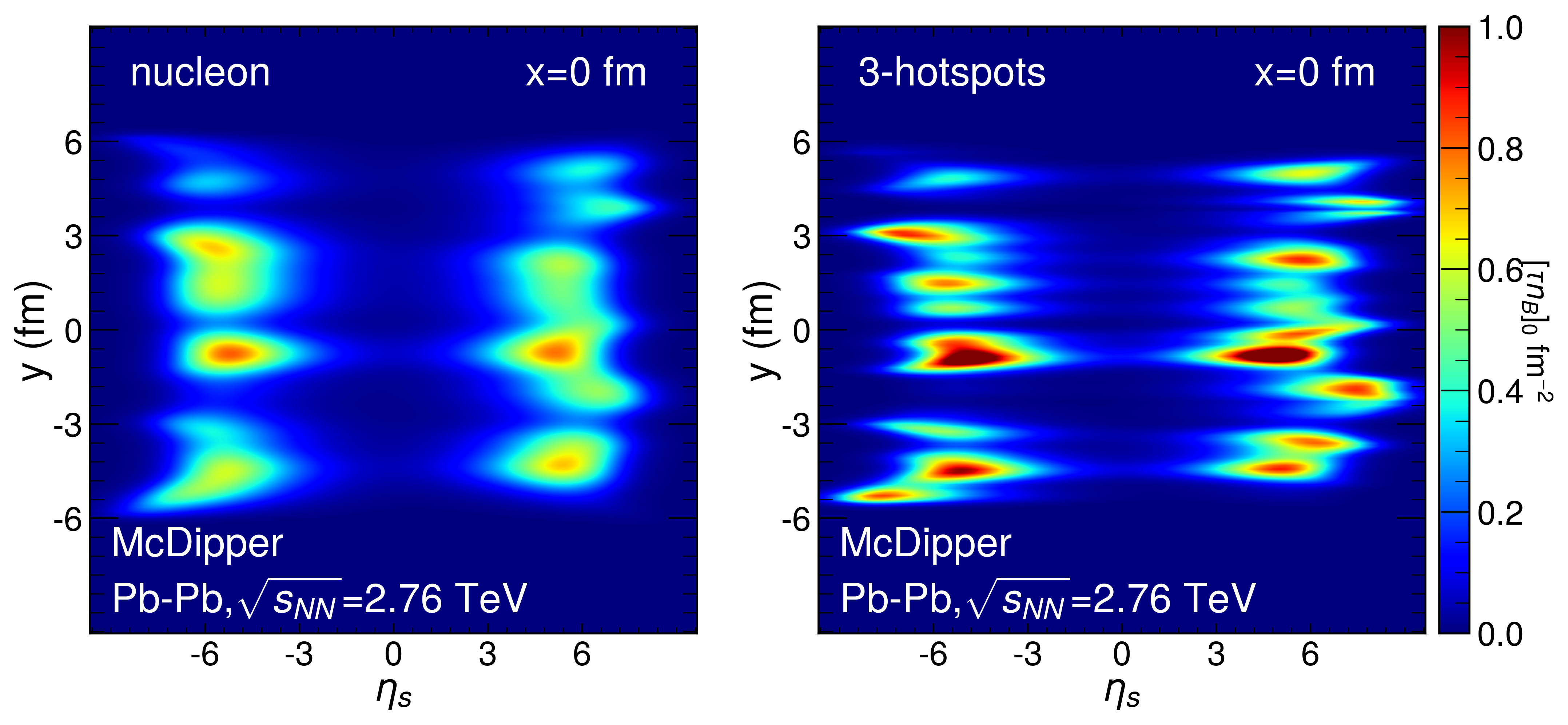}
    \includegraphics[width=\linewidth]{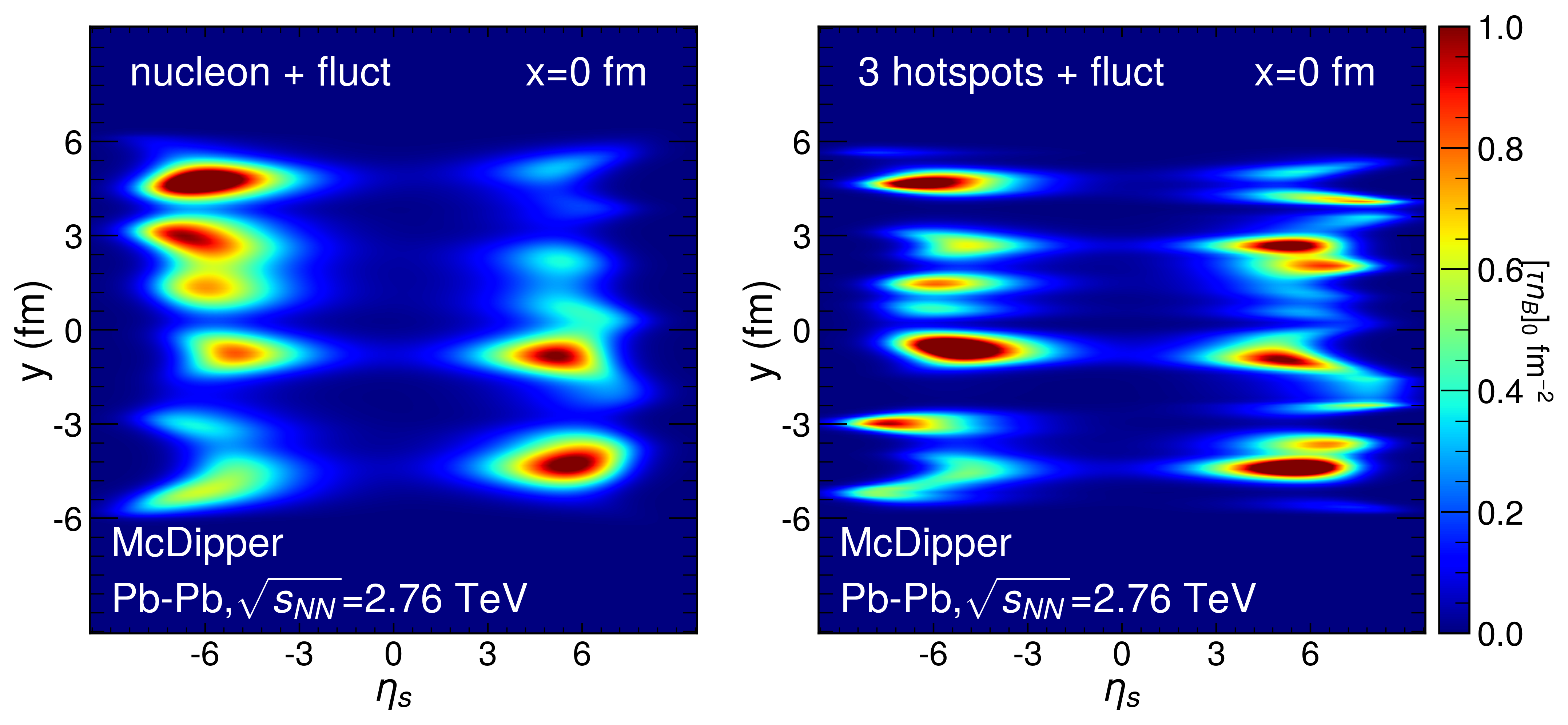}
    \caption{The initial net-baryon number density distributions of different classes at x=0 fm for a sampled Pb+Pb collision at $\sqrt{s_{NN}}$=2.76 TeV and impact parameter b=3.42 fm.}
    \label{fig:ini-nb-long}
\end{figure}

In Fig. \ref{fig:ini-ed-long} and \ref{fig:ini-nb-long}, we present the initial energy density and net-baryon number distributions in longitudinal direction at x=0 fm with four different classes of initial conditions.
Similar to the transverse energy density distribution, one can see that the inclusion of 3 hotspots in each nucleon causes the longitudinal pattern to be divided into thinner slices which may influence longitudinal behavior of subsequent expansion. On the other hand, we can clearly see that the initial net-baryon deposition is always located in the forward or backward rapidity regions, and the sub-nucleonic fluctuations have similar effects on baryon stopping as on energy deposition.

\section{Elliptic flow in event plane method}\label{app-v2}
\begin{figure*}
    \centering
    \includegraphics[width=0.9\linewidth]{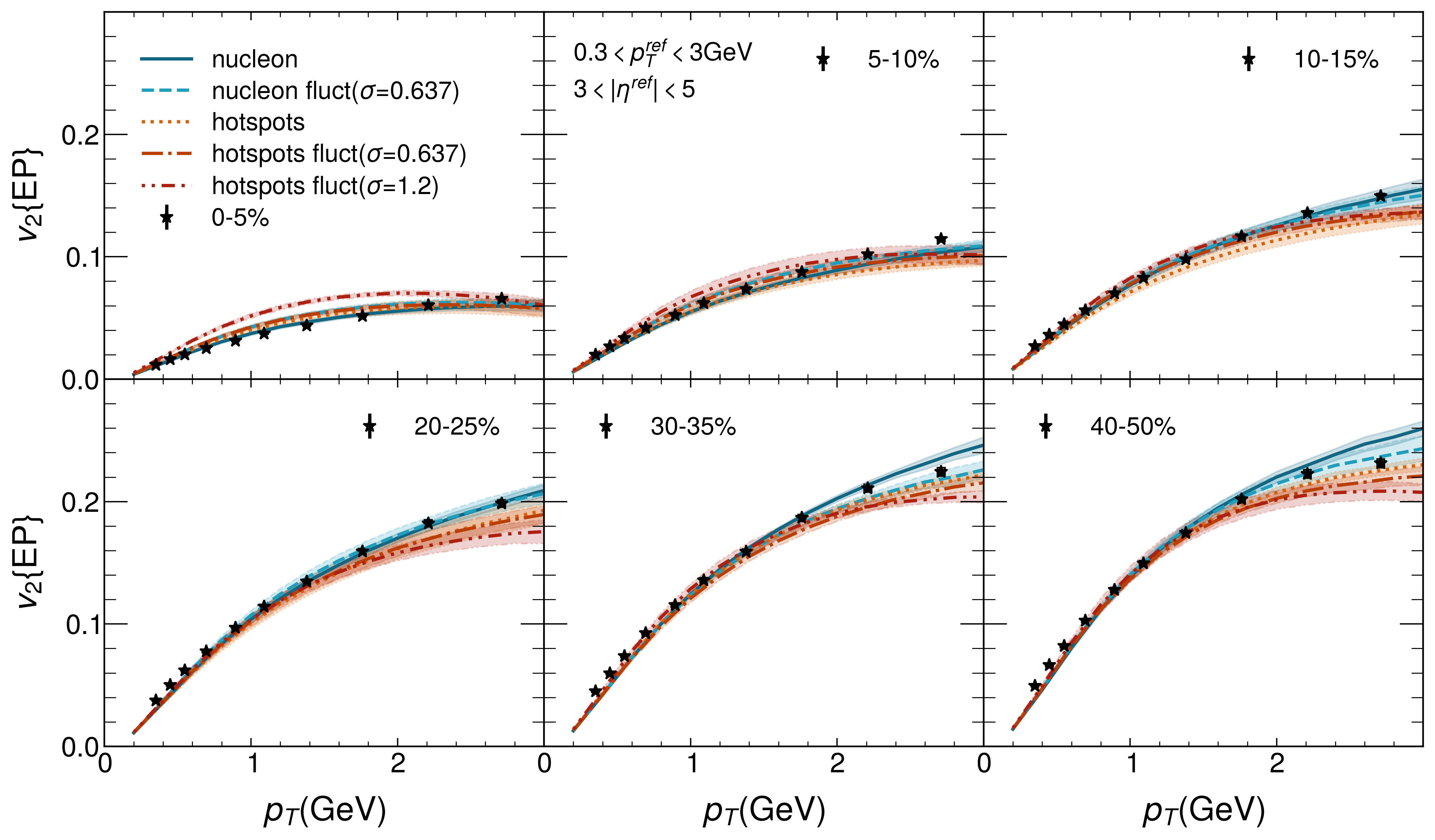}
    \caption{The elliptic flow, $v_2\{\rm EP\}$(curves) calculated with event plane method, for charged particles  as a function of transverse momentum $p_T$  in Pb+Pb collisions at $\sqrt{s_{NN}}$ = 2.76 TeV. 
    The data(points) are taken from CMS\cite{CMS:2012zex}.
    The shaded areas around each line represent the statistical uncertainty of the simulation results.}
    \label{fig:v2pt-ep}
\end{figure*}

\begin{figure*}
    \centering
    \includegraphics[width=0.9\linewidth]{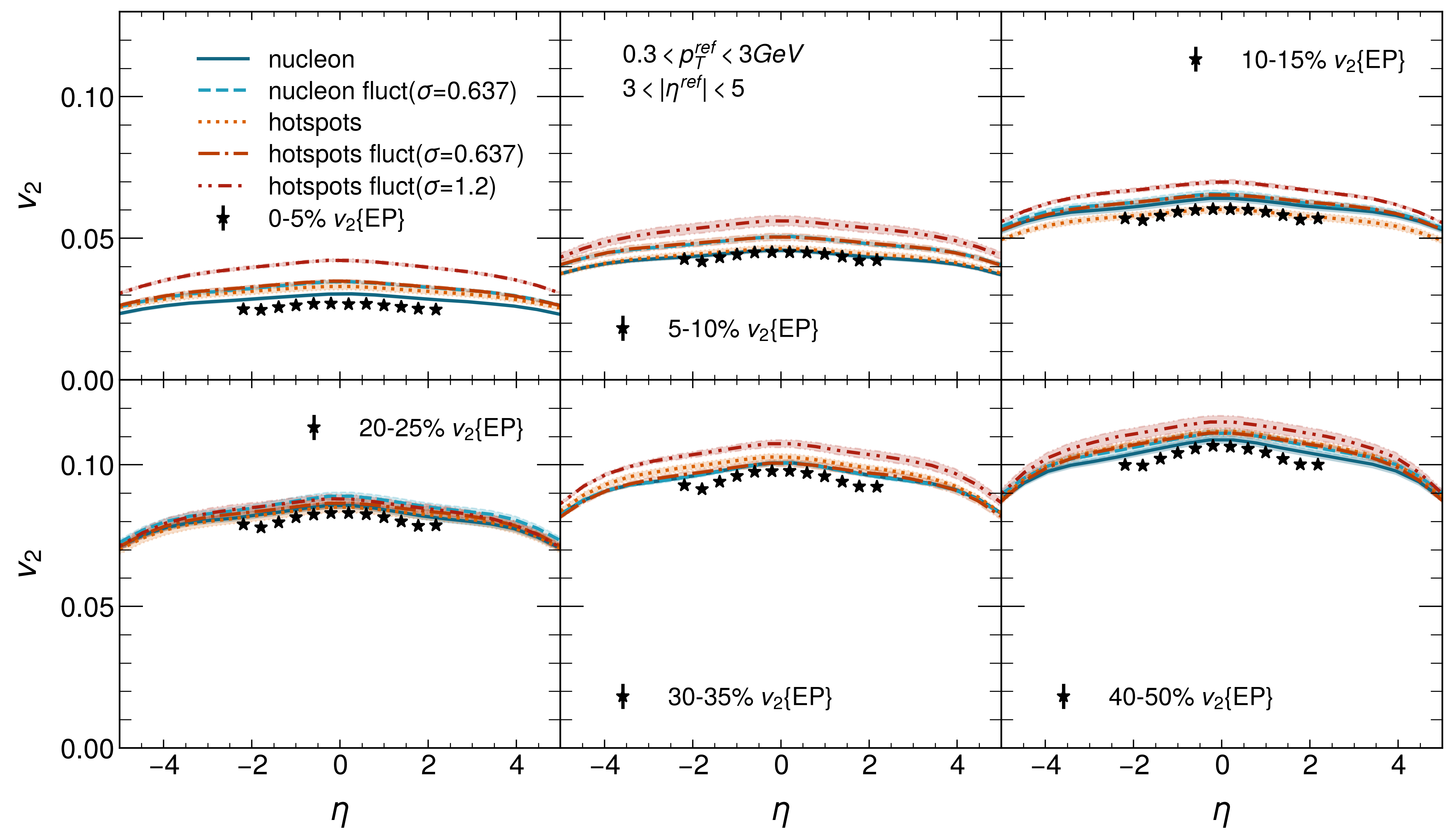}
    \caption{The rapidity dependence of the $v_2(\eta)$ calculated with event plane method, compared to CMS\cite{CMS:2012zex} data. Both data and the calculation are for 0.3<$p_T$<3 GeV and use reference particles in (3<|$\eta$|<5).
    The shaded areas around each line represent the statistical uncertainty of the simulation results.}
    \label{fig:v2eta-cms-ep}
\end{figure*}

Some experimental results on elliptic flow are calculated with event plane method\cite{CMS:2012zex, Poskanzer:1998yz},

\begin{equation}
v_n\{E P\}=
\frac{\left\langle\left\{\cos n\left(\phi-\tilde{\Phi}_{n A}\right)\right\}\right\rangle}
{\sqrt{\left\langle\cos n\left(\tilde{\Phi}_{n A}-\tilde{\Phi}_{n B}\right)\right\rangle}},
\end{equation}

where A and B denote two sub-events from different rapidity ranges. The event plane $\tilde{\Phi}_{n A(B)}$ are defined as $$\tilde{\Phi}_{n}=\frac{1}{n}\arctan \frac{\langle\sin(n\phi)\rangle}{\langle\cos(n\phi)\rangle}.$$
The $p_T$-dependence and $\eta$-dependence of elliptic flow calculated with event plane method are shown in Fig. \ref{fig:v2pt-ep} and Fig. \ref{fig:v2eta-cms-ep}. They confirmed the previous conclusion drawn from the flow measurements using the multi-particle cumulant method in Sec. \ref{subsec-v2}.

\section{Hydrodynamic tuning}\label{app-hydro-tune}
\begin{figure*}[b]
  \centering
  \begin{subfigure}[b]{0.49\textwidth}
    \includegraphics[width=\textwidth]{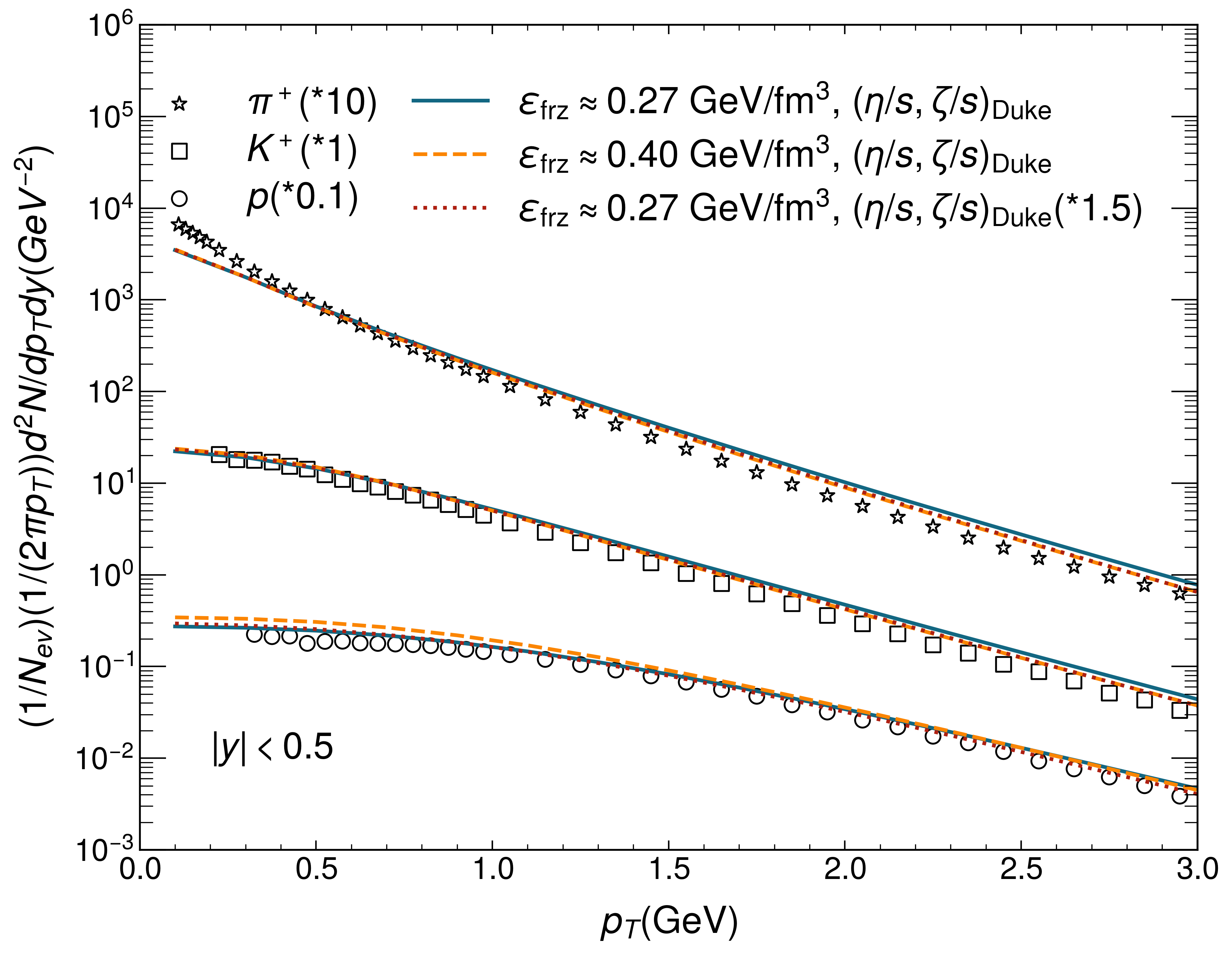}
  \end{subfigure}
  \hfill
  \begin{subfigure}[b]{0.49\textwidth}
    \includegraphics[width=\textwidth]{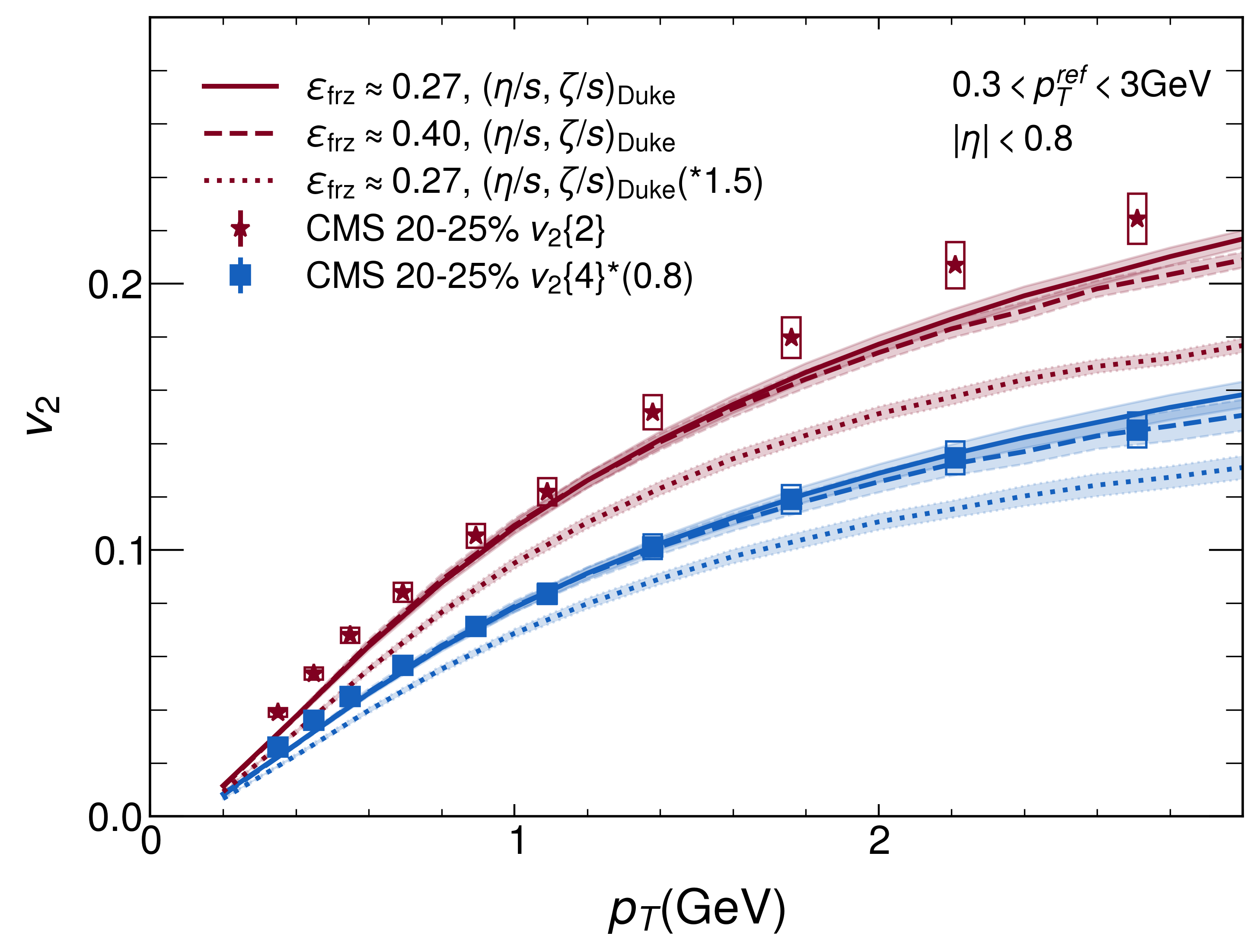}
  \end{subfigure}
  \caption{
    The effects of freeze-out energy density($\varepsilon_{\rm frz}$) and viscosity on transverse momentum spectra for identified particles(left panel) and elliptic flow for charged particles as a function of transverse momentum $p_T$(right panel) for Pb+Pb collisions at $\sqrt{s_{NN}}$=2.76 TeV.
    The shaded areas represent the statistical uncertainty.
  }
  \label{fig:prd-pT-v2-pT}
\end{figure*}

\begin{figure*}[b]
  \centering
  \begin{subfigure}[b]{0.49\textwidth}
    \includegraphics[width=\textwidth]{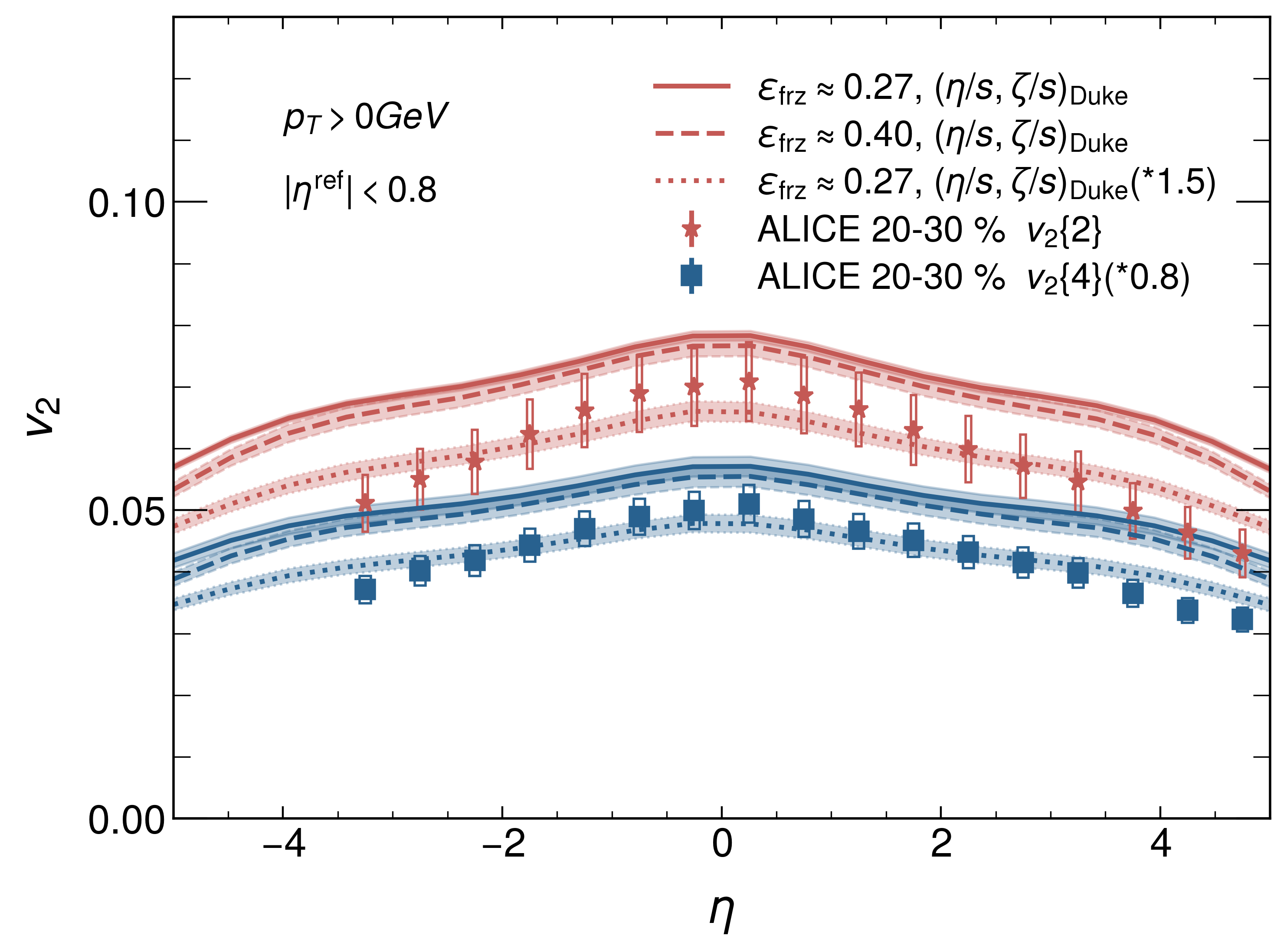}
    %\label{fig:prd-v2-alice}
  \end{subfigure}
  \hfill
  \begin{subfigure}[b]{0.49\textwidth}
    \includegraphics[width=\textwidth]{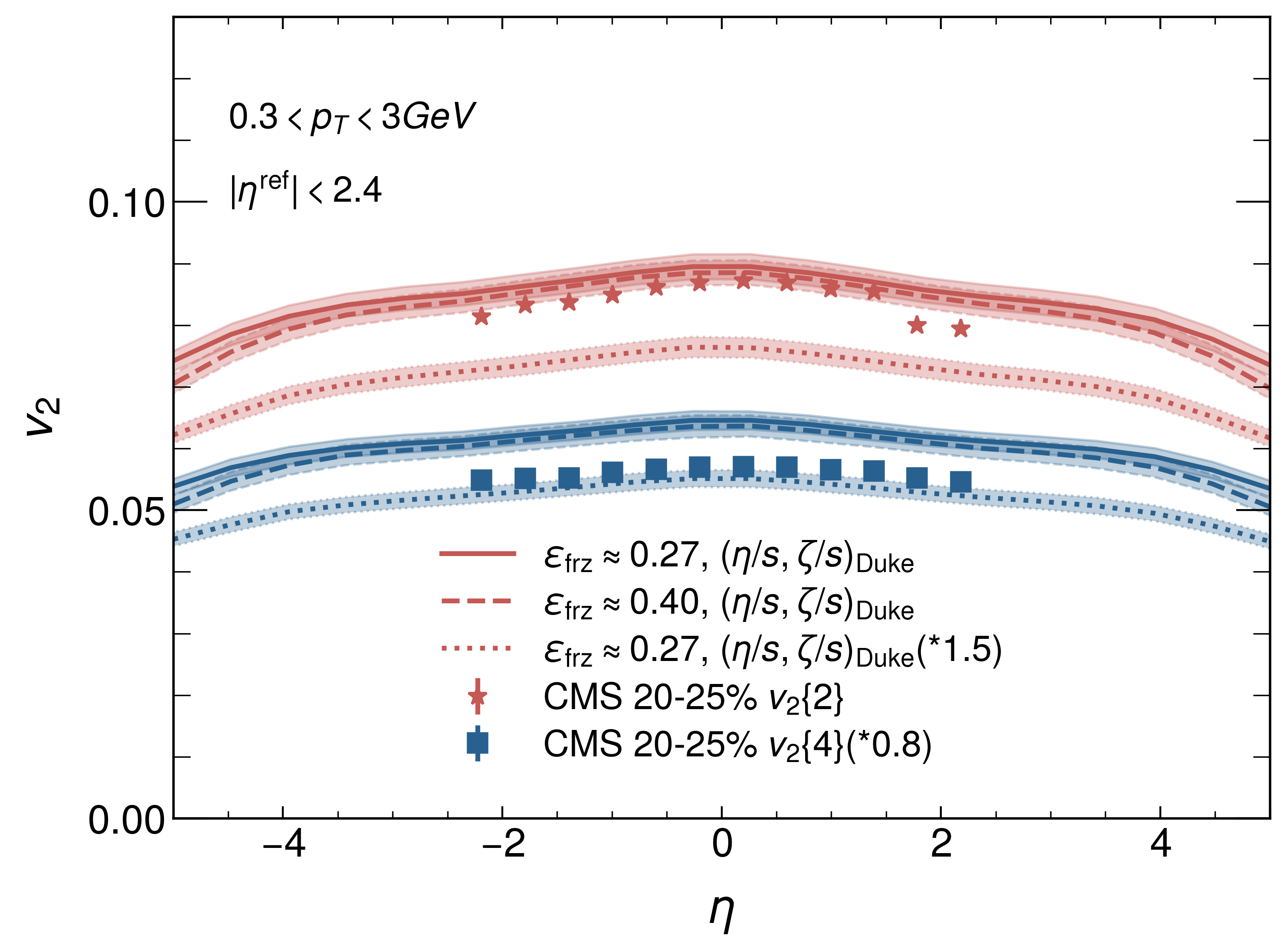}
    %\label{fig:prd-v2-cms}
  \end{subfigure}
  \caption{The effects of freeze-out energy density($\varepsilon_{\rm frz}$) and viscosity on rapidity dependence of the momentum anisotropies $v_2(\eta)$.
    The left panel is comparison to ALICE results while the right panel is for comparison to CMS measurements.
    The shaded areas represent the statistical uncertainty.
    }
   \label{fig:prd-v2-eta}
\end{figure*}

Even though this paper does not aim to describe experimental data, we perform a minimal hydrodynamic tuning to demonstrate the potential of our framework.
Due to the absence of a hadronic cascade in the current setup, the freeze-out energy density is expected to influence final observables, such as the transverse momentum spectra of identified particles. Therefore, we vary the freeze-out energy density from 0.27 to 0.4 GeV/fm$^3$. 
In addition, the shear and bulk viscosity have not yet been fully constrained.
The values used in this work are taken from the Bayesian analysis by the Duke group~\cite{Bernhard:2016tnd}, however, such results may depend on specific model assumptions. So we simply increase them by a factor of 1.5 to suppress the flow results and $p_T$ spectra.
As shown in the left panel of Fig.~\ref{fig:prd-pT-v2-pT}, a higher freeze-out energy density or larger viscosity pushes more particles to low-$p_T$ region.
On the other hand, from Fig.~\ref{fig:prd-v2-eta}, we observe that larger viscosity significantly suppresses elliptic flow while a higher freeze-out energy density only leads to a slight suppression.
Furthermore, the right panel of Fig.~\ref{fig:prd-pT-v2-pT} indicates that the suppression of the integrated elliptic flow originates from its $p_T$ dependence. Overall, careful tuning of parameters is necessary to simultaneously describe various observables. In this context, it would also be desirable to include SMASH as an afterburner, since $p_T$-differential observables in particular are expected to be sensitive to hadronic rescattering effects~\cite{Ryu:2015vwa, Ryu:2017qzn, Teaney:2000cw, Bass:2000ib}. The results presented in this paper are intended to highlight the potential of our approach, and we prefer to postpone further tuning until after the afterburner is incorporated.

%\begin{figure}[h]
%    \centering
%    \includegraphics[width=\linewidth]{PRD_v2_vs_eta_cms_ep.png}
%    \caption{The effects of freeze-out energy density($\varepsilon_{\rm frz}$) and viscosity on rapidity dependence of the $v_2(\eta)$ calculated with event plane method. The shaded areas represent the statistical uncertainty.}
%    \label{fig:prd-v2-3}
%\end{figure}

\end{document}